\documentclass[twocolumn,epjc3]{svjour3}

\input{preamble}

\tikzstyle{Black circle}=[fill=black, draw=black, shape=circle]
\tikzstyle{Red circle}=[fill=red, draw=red, shape=circle]
\tikzstyle{Green circle}=[fill=green, draw=green, shape=circle]
\tikzstyle{Blue circle}=[fill=blue, draw=blue, shape=circle]
\tikzstyle{orange hue}=[fill={rgb,255: red,255; green,193; blue,6}, draw=none, shape=circle, tikzit fill={rgb,255: red,255; green,193; blue,6}]
\tikzstyle{write yellow}=[fill=none, draw=none, shape=circle]

\tikzstyle{Red edge}=[-, fill=red, draw=red, line width=0.5mm]
\tikzstyle{Blue line}=[-, fill=blue, draw=blue, line width=0.5mm]
\tikzstyle{red arrow}=[fill=red, draw=red, <->, line width=0.5mm]
\tikzstyle{Blue arrow}=[fill=blue, draw=blue, <->, line width=0.5mm]
\tikzstyle{Orange line}=[-, fill={rgb,255: red,255; green,216; blue,21}, draw={rgb,255: red,255; green,216; blue,21},]
\tikzstyle{new edge style 0}=[fill={rgb,255: red,255; green,216; blue,21}, draw={rgb,255: red,255; green,216; blue,21}, ->, line width=0.5mm]

\tikzstyle{Red edge2}=[-, fill=red, draw=red, line width=0.4mm]
\tikzstyle{Blue line2}=[-, fill=blue, draw=blue, line width=0.4mm]
\tikzstyle{red arrow2}=[fill=red, draw=red, <->, line width=0.4mm]
\tikzstyle{Blue arrow2}=[fill=blue, draw=blue, <->, line width=0.4mm]
\tikzstyle{Orange line2}=[-, draw={rgb,255: red,255; green,216; blue,21}, line width=0.4mm]
\tikzstyle{new edge style 02}=[draw={rgb,255: red,255; green,216; blue,21}, ->, line width=0.4mm]
\tikzstyle{axis2}=[->, draw={rgb,255: red,64; green,64; blue,64}, line width=0.4mm]
\tikzstyle{Pink line2}=[-, draw={rgb,255: red,255; green,171; blue,251}, line width=0.4mm]
\tikzstyle{Red dashed2}=[fill=red, draw=red, dashed, ->, line width=0.4mm]
\tikzstyle{dashed pink2}=[draw={rgb,255: red,255; green,171; blue,251}, dashed, ->, line width=0.4mm]

\journalname{Eur. Phys. J. C}

\begin{document}

\title{Testing T2K's Bayesian constraints with priors in alternate parameterisations}

\institute{ University Autonoma Madrid, Department of Theoretical Physics, 28049 Madrid, Spain \label{INSTHD} \and\pagebreak[0] University of British Columbia, Department of Physics and Astronomy, Vancouver, British Columbia, Canada \label{INSTD} \and\pagebreak[0] Boston University, Department of Physics, Boston, Massachusetts, U.S.A. \label{INSTFE} \and\pagebreak[0] University of California, Irvine, Department of Physics and Astronomy, Irvine, California, U.S.A. \label{INSTGA} \and\pagebreak[0] IRFU, CEA, Universit\'e Paris-Saclay, F-91191 Gif-sur-Yvette, France \label{INSTI} \and\pagebreak[0] University of Colorado at Boulder, Department of Physics, Boulder, Colorado, U.S.A. \label{INSTGB} \and\pagebreak[0] Colorado State University, Department of Physics, Fort Collins, Colorado, U.S.A. \label{INSTFG} \and\pagebreak[0] Duke University, Department of Physics, Durham, North Carolina, U.S.A. \label{INSTFH} \and\pagebreak[0] E\"{o}tv\"{o}s Lor\'{a}nd University, Department of Atomic Physics, Budapest, Hungary \label{INSTJA} \and\pagebreak[0] ETH Zurich, Institute for Particle Physics and Astrophysics, Zurich, Switzerland \label{INSTEF} \and\pagebreak[0] University of Science, Vietnam National University, Hanoi, Vietnam \label{INSTIG} \and\pagebreak[0] CERN European Organization for Nuclear Research, CH-1211 Gen\'eve 23, Switzerland \label{INSTIE} \and\pagebreak[0] University of Geneva, Section de Physique, DPNC, Geneva, Switzerland \label{INSTEG} \and\pagebreak[0] University of Glasgow, School of Physics and Astronomy, Glasgow, United Kingdom \label{INSTHJ} \and\pagebreak[0] Ghent University, Department of Physics and Astronomy, Proeftuinstraat 86, B-9000 Gent, Belgium \label{INSTJG} \and\pagebreak[0] H. Niewodniczanski Institute of Nuclear Physics PAN, Cracow, Poland \label{INSTDG} \and\pagebreak[0] High Energy Accelerator Research Organization (KEK), Tsukuba, Ibaraki, Japan \label{INSTCB} \and\pagebreak[0] University of Houston, Department of Physics, Houston, Texas, U.S.A. \label{INSTIB} \and\pagebreak[0] Institut de Fisica d'Altes Energies (IFAE) - The Barcelona Institute of Science and Technology, Campus UAB, Bellaterra (Barcelona) Spain \label{INSTED} \and\pagebreak[0] Institut f\"ur Physik, Johannes Gutenberg-Universit\"at Mainz, Staudingerweg 7, 55128 Mainz, Germany \label{INSTJC} \and\pagebreak[0] IFIC (CSIC \& University of Valencia), Valencia, Spain \label{INSTEC} \and\pagebreak[0] Institute For Interdisciplinary Research in Science and Education (IFIRSE), ICISE, Quy Nhon, Vietnam \label{INSTHH} \and\pagebreak[0] Imperial College London, Department of Physics, London, United Kingdom \label{INSTEI} \and\pagebreak[0] INFN Sezione di Bari and Universit\`a e Politecnico di Bari, Dipartimento Interuniversitario di Fisica, Bari, Italy \label{INSTGF} \and\pagebreak[0] INFN Sezione di Napoli and Universit\`a di Napoli, Dipartimento di Fisica, Napoli, Italy \label{INSTBE} \and\pagebreak[0] INFN Sezione di Padova and Universit\`a di Padova, Dipartimento di Fisica, Padova, Italy \label{INSTBF} \and\pagebreak[0] INFN Sezione di Roma and Universit\`a di Roma ``La Sapienza'', Roma, Italy \label{INSTBD} \and\pagebreak[0] Institute for Nuclear Research of the Russian Academy of Sciences, Moscow, Russia \label{INSTEB} \and\pagebreak[0] International Centre of Physics, Institute of Physics (IOP), Vietnam Academy of Science and Technology (VAST), 10 Dao Tan, Ba Dinh, Hanoi, Vietnam \label{INSTHI} \and\pagebreak[0] ILANCE, CNRS – University of Tokyo International Research Laboratory, Kashiwa, Chiba 277-8582, Japan \label{INSTJD} \and\pagebreak[0] Kavli Institute for the Physics and Mathematics of the Universe (WPI), The University of Tokyo Institutes for Advanced Study, University of Tokyo, Kashiwa, Chiba, Japan \label{INSTHA} \and\pagebreak[0] Keio University, Department of Physics, Kanagawa, Japan \label{INSTID} \and\pagebreak[0] King's College London, Department of Physics, Strand, London WC2R 2LS, United Kingdom \label{INSTIF} \and\pagebreak[0] Kobe University, Kobe, Japan \label{INSTCC} \and\pagebreak[0] Kyoto University, Department of Physics, Kyoto, Japan \label{INSTCD} \and\pagebreak[0] Lancaster University, Physics Department, Lancaster, United Kingdom \label{INSTEJ} \and\pagebreak[0] Lawrence Berkeley National Laboratory, Berkeley, CA 94720, USA \label{INSTII} \and\pagebreak[0] Ecole Polytechnique, IN2P3-CNRS, Laboratoire Leprince-Ringuet, Palaiseau, France \label{INSTBA} \and\pagebreak[0] University of Liverpool, Department of Physics, Liverpool, United Kingdom \label{INSTFC} \and\pagebreak[0] Louisiana State University, Department of Physics and Astronomy, Baton Rouge, Louisiana, U.S.A. \label{INSTFI} \and\pagebreak[0] Joint Institute for Nuclear Research, Dubna, Moscow Region, Russia \label{INSTIH} \and\pagebreak[0] Michigan State University, Department of Physics and Astronomy,  East Lansing, Michigan, U.S.A. \label{INSTHB} \and\pagebreak[0] Miyagi University of Education, Department of Physics, Sendai, Japan \label{INSTCE} \and\pagebreak[0] National Centre for Nuclear Research, Warsaw, Poland \label{INSTDF} \and\pagebreak[0] State University of New York at Stony Brook, Department of Physics and Astronomy, Stony Brook, New York, U.S.A. \label{INSTFJ} \and\pagebreak[0] Okayama University, Department of Physics, Okayama, Japan \label{INSTGJ} \and\pagebreak[0] Osaka Metropolitan University, Department of Physics, Osaka, Japan \label{INSTCF} \and\pagebreak[0] Oxford University, Department of Physics, Oxford, United Kingdom \label{INSTGG} \and\pagebreak[0] University of Pennsylvania, Department of Physics and Astronomy,  Philadelphia, PA, 19104, USA. \label{INSTIC} \and\pagebreak[0] University of Pittsburgh, Department of Physics and Astronomy, Pittsburgh, Pennsylvania, U.S.A. \label{INSTGC} \and\pagebreak[0] Queen Mary University of London, School of Physics and Astronomy, London, United Kingdom \label{INSTFA} \and\pagebreak[0] University of Regina, Department of Physics, Regina, Saskatchewan, Canada \label{INSTE} \and\pagebreak[0] University of Rochester, Department of Physics and Astronomy, Rochester, New York, U.S.A. \label{INSTGD} \and\pagebreak[0] Royal Holloway University of London, Department of Physics, Egham, Surrey, United Kingdom \label{INSTHC} \and\pagebreak[0] RWTH Aachen University, III. Physikalisches Institut, Aachen, Germany \label{INSTBC} \and\pagebreak[0] School of Physics and Astronomy, University of Minnesota, Minneapolis, Minnesota, U.S.A. \label{INSTJF} \and\pagebreak[0] Departamento de F\'isica At\'omica, Molecular y Nuclear, Universidad de Sevilla, 41080 Sevilla, Spain \label{INSTJB} \and\pagebreak[0] University of Sheffield, Department of Physics and Astronomy, Sheffield, United Kingdom \label{INSTFB} \and\pagebreak[0] University of Silesia, Institute of Physics, Katowice, Poland \label{INSTDI} \and\pagebreak[0] SLAC National Accelerator Laboratory, Stanford University, Menlo Park, California, U.S.A. \label{INSTIA} \and\pagebreak[0] Sorbonne Universit\'e, Universit\'e Paris Diderot, CNRS/IN2P3, Laboratoire de Physique Nucl\'eaire et de Hautes Energies (LPNHE), Paris, France \label{INSTBB} \and\pagebreak[0] South Dakota School of Mines and Technology, 501 East Saint Joseph Street, Rapid City, SD 57701, United States \label{INSTJE} \and\pagebreak[0] STFC, Rutherford Appleton Laboratory, Harwell Oxford,  and  Daresbury Laboratory, Warrington, United Kingdom \label{INSTEH} \and\pagebreak[0] University of Tokyo, Department of Physics, Tokyo, Japan \label{INSTCH} \and\pagebreak[0] University of Tokyo, Institute for Cosmic Ray Research, Kamioka Observatory, Kamioka, Japan \label{INSTBJ} \and\pagebreak[0] University of Tokyo, Institute for Cosmic Ray Research, Research Center for Cosmic Neutrinos, Kashiwa, Japan \label{INSTCG} \and\pagebreak[0] Tokyo Institute of Technology, Department of Physics, Tokyo, Japan \label{INSTHF} \and\pagebreak[0] Tokyo Metropolitan University, Department of Physics, Tokyo, Japan \label{INSTGI} \and\pagebreak[0] Tokyo University of Science, Faculty of Science and Technology, Department of Physics, Noda, Chiba, Japan \label{INSTHG} \and\pagebreak[0] University of Toronto, Department of Physics, Toronto, Ontario, Canada \label{INSTF} \and\pagebreak[0] TRIUMF, Vancouver, British Columbia, Canada \label{INSTB} \and\pagebreak[0] University of Toyama, Department of Physics, Toyama, Japan \label{INSTJH} \and\pagebreak[0] University of Warsaw, Faculty of Physics, Warsaw, Poland \label{INSTDJ} \and\pagebreak[0] Warsaw University of Technology, Institute of Radioelectronics and Multimedia Technology, Warsaw, Poland \label{INSTDH} \and\pagebreak[0] Tohoku University, Faculty of Science, Department of Physics, Miyagi, Japan \label{INSTIJ} \and\pagebreak[0] University of Warwick, Department of Physics, Coventry, United Kingdom \label{INSTFD} \and\pagebreak[0] University of Winnipeg, Department of Physics, Winnipeg, Manitoba, Canada \label{INSTGH} \and\pagebreak[0] Wroclaw University, Faculty of Physics and Astronomy, Wroclaw, Poland \label{INSTEA} \and\pagebreak[0] Yokohama National University, Department of Physics, Yokohama, Japan \label{INSTHE} \and\pagebreak[0] York University, Department of Physics and Astronomy, Toronto, Ontario, Canada \label{INSTH}}
\thankstext{thanks0}{also at Universit\'e Paris-Saclay} \thankstext{thanks2}{also at J-PARC, Tokai, Japan} \thankstext{thanks3}{affiliated member at Kavli IPMU (WPI), the University of Tokyo, Japan} \thankstext{thanks4}{also at Moscow Institute of Physics and Technology (MIPT), Moscow region, Russia and National Research Nuclear University "MEPhI", Moscow, Russia} \thankstext{thanks5}{also at IPSA-DRII, France} \thankstext{thanks6}{also at the Graduate University of Science and Technology, Vietnam Academy of Science and Technology} \thankstext{thanks7}{also at JINR, Dubna, Russia} \thankstext{thanks8}{also at Nambu Yoichiro Institute of Theoretical and Experimental Physics (NITEP)} \thankstext{thanks9}{also at BMCC/CUNY, Science Department, New York, New York, U.S.A.} \thankstext{thanks10}{also at Departament de Fisica de la Universitat Autonoma de Barcelona, Barcelona, Spain.}
\author{The T2K Collaboration: K.\,Abe\thanksref{INSTBJ}, S.\,Abe\thanksref{INSTBJ}, R.\,Akutsu\thanksref{INSTCB}, H.\,Alarakia-Charles\thanksref{INSTEJ}, Y.I.\,Alj Hakim\thanksref{INSTFB}, S.\,Alonso Monsalve\thanksref{INSTEF}, L.\,Anthony\thanksref{INSTEI}, S.\,Aoki\thanksref{INSTCC}, K.A.\,Apte\thanksref{INSTEI}, T.\,Arai\thanksref{INSTCH}
% ...
}

\date{Received: 02-Jul-2025}
\onecolumn
\maketitle

\begin{abstract}
Bayesian analysis results require a choice of prior distribution. In long-baseline neutrino oscillation physics, the usual parameterisation of the mixing matrix induces a prior that privileges certain neutrino mass and flavour state symmetries. Here we study the effect of privileging alternate symmetries on the results of the T2K experiment. We find that constraints on the level of CP violation (as given by the Jarlskog invariant) are robust under the choices of prior considered in the analysis. On the other hand, the degree of octant preference for the atmospheric angle depends on which symmetry has been privileged.

\end{abstract}

\twocolumn
\section{Introduction}
Bayesian analyses have become powerful tools in accelerator long-baseline neutrino oscillation measurements~\cite{mach3}~\cite{aria}, due to their flexibility in incorporating non-Gaussian features, highly degenerate parameters, and post-analysis interpretation of results. However, in these Bayesian analyses, the choice of prior distribution may impact the results, and understanding this effect of prior choice is important to interpreting them~\cite{priors}.

\sloppy Neutrino oscillations are typically described using a unitary mixing matrix, $U_{PMNS}$, called the Ponte\-corvo–Maki–Naka\-gawa–Sa\-kata (PMNS) matrix~\cite{Pontecorvo}\cite{MakiNakagawaSakata}, using a common parameterisation, described in Section~\ref{sec:parameterisation}. The physical manifestation of neutrino oscillations depends only on the moduli of the matrix elements, yet the parameters in the common parameterisation are not closely related to the physical observables $|U_{\alpha i}|^2$. This fact means that commonly used simple priors may not fully reflect the underlying physics or potential symmetries of the matrix.

This work explores the flavour symmetry biases induced by uniform priors on the parameters of the standard PMNS parameterisation, and investigates alternatives that bring out the flavour and mass symmetry preferences intrinsic to other parameterisations. The new priors are applied to T2K's 2022 neutrino oscillation results~\cite{oa2023} to quantify the robustness of its constraints. Section~\ref{sec:parameterisation} develops a framework for finding useful alternate parameterisation, and Section~\ref{sec:choice} an interpretation of the parameterisations used in this analysis. Section~\ref{sec:MCMC} discusses the technique used to implement the new priors in T2K's analysis and the uncertainties that arise with it. Finally, Section~\ref{sec:results} reports the variations in T2K's  constraints induced by choosing a different prior.

\section{Parameterisations of the leptonic mixing matrix}
\label{sec:parameterisation}
The standard parameterisation of the PMNS matrix, used in the Particle Data Group's (PDG) summary~\cite{pdg}, was inherited from the quark sector~\cite{originalCKMparams} and proved useful for describing early results in neutrino oscillation~\cite{SNOfirstResults}~\cite{SKfirst}. It represents the mixing matrix using three Tait-Bryan rotation angles~\cite{taitBryan} $(\theta_{12}, \theta_{23},\theta_{13})$ and a complex phase $\delta_{CP}$ under the following construction \cite{pdgForm}:

\begin{equation}
    U_{PMNS} \equiv U_{R} = R_{23}\Gamma_\delta^\dagger R_{13} \Gamma_\delta R_{12} ,
    \label{eqn:PDG}
\end{equation}
where, using the abbreviations $s_{ij}\equiv\sin\theta_{ij}$ and $c_{ij}\equiv\cos\theta_{ij}$,

\begin{align}
    & R_{23} \equiv \begin{pmatrix}
        1 & 0 & 0 \\
        0 & c_{23} & s_{23} \\
        0 & -s_{23} & c_{23}
    \end{pmatrix} \;\;\;\;
    R_{13} \equiv \begin{pmatrix}
        c_{13} & 0 & s_{13} \\
        0 & 1 & 0 \\
        -s_{13} & 0 & c_{13}
    \end{pmatrix} \nonumber \\ \\
    & R_{12} \equiv \begin{pmatrix}
        c_{12} & s_{12} & 0 \\
        -s_{12} & c_{12} & 0 \\
        0 & 0 & 1
    \end{pmatrix} \;\;\;\;
    \Gamma_\delta \equiv \begin{pmatrix}
        e^{i\delta_{CP}} & 0\;\; & 0 \\
        0 & 1\;\; & 0 \\
        0 & 0\;\; & 1
    \end{pmatrix}. \nonumber
    \label{eqn:rotationMatrices}
\end{align}

Since $U_{PMNS}$ transforms from the mass to the flavour eigenstates, one can see that $R_{12}$ acts directly on the mass basis and is therefore a rotation of the $(\nu_1\nu_2)$ plane of mass states. Similarly, $R_{23}$ acts on the flavour basis and is a rotation of the $(\nu_\mu\nu_\tau)$ plane of flavour states. Finally, $\Gamma_\delta R_{13} \Gamma_\delta^\dagger$ is a rotation around the ($\nu_3\nu_e$) plane, involving both flavour and mass states.

This construction became the standard in neutrino oscillation analyses because when using it, the smallness of $|U_{e3}|$ and the hierarchical structure of the masses ($\Delta m^2_{21} \ll \Delta m^2_{32}$) conspire to make the expressions for solar and atmospheric mixing surprisingly simple. For atmospheric neutrino energies and oscillation distances, the $\Delta m^2_{32}$ contribution dominates, and so we can approximate the oscillation probabilities as $\nu_\mu \leftrightarrow \nu_\tau$ mixing with a small contamination from $\nu_e$. In the MSW limit~\cite{MSW1}\cite{MSW2}, solar experiments become a measurement of the $\nu_2$ component projected onto $\nu_e$. Moreover, $|U_{e3}|$ being small, $\nu_e$ survival can be studied as $\nu_1 \leftrightarrow\nu_2$  mixing with a small correction from a weakly mixed $\nu_3$ state. 

In the canonical (or standard) parameterisation\footnote{Often referred to as the PDG parameterisation~\cite{PDGscheme}.} the angle $\theta_{23}^\text{PDG}$ (which defines $R_{23}$) mixes $\nu_\mu$ with $\nu_\tau$ and is a natural extension of the mixing angle in the 2-flavour atmospheric approximation. In the solar sector, the MSW resonance is very close to a direct measurement of the inner product $\langle\nu_e, \nu_2\rangle$ $(=|U_{e2}|)$, so a parameterisation that has $U_{e2}$ as the simple element, written as $\sin\theta_{13}e^{-i\delta_{CP}}$, would be most optimal for solar experiments\footnote{We can identify the $\nu_\mu\nu_\tau/\nu_1\nu_3$ parameterisation presented in section \ref{sec:parameterisation} as the matrix in question.}. Instead, $|U_{e3}|$ is small enough to allow for the $\nu_1,\nu_2$ two-flavour approximation and makes the standard scheme convenient.

In general, one is free to choose the rotation axes for these Tait-Bryan rotations, and any set of perpendicular rotations will give rise to a valid parameterisation of the PMNS matrix. Although we have infinitely many choices, the only bases we have a reason to work with are the mass and flavour bases; therefore, we restrict ourselves to working with rotations defined around those. More formally, the infinite choices can be accessed by introducing additional U(3) rotations encoded in matrices $X_1,X_2$ to shift the axes to their desired positions
\begin{equation}
    U_{PMNS} \equiv U_{X_1X_2} = X_1U_{R}X_2
    \label{eqn:genericPars}
\end{equation}

As long as $X_i$ are not the identity, identical values for the angles $\theta_{ij}$ will lead to different matrices ($U_{X_1X_2}\neq U_{R}$). If we want to describe the same neutrino mixing using the $X$ and PDG forms of the PMNS matrix we must find two sets of mixing parameters $\theta_{ij}^{X_1X_2}, \delta_{CP}^{X_1X_2}$ and $\theta_{ij}^{PDG}, \delta_{CP}^{PDG}$ that fulfil $|U_{X_1X_2}|_{\alpha k}= |U_{PDG}|_{\alpha k}$. That is, each choice of $X_i$ gives rise to a new Tait-Bryan parameterisation which redefines the meaning of the mixing parameters.

We have discussed the origin of the canonical parameterisation in neutrino mixing and shown that there are many choices of Tait-Bryan parameterisations. Under no assumptions of hierarchy, these are all equivalent. In the next section, we motivate choices of particular Tait-Bryan parameterisations for oscillation analysis.

\section{Choosing a parameterisation for Bayesian long-baseline oscillation analysis}
\label{sec:choice}

\sloppy Neutrino oscillation analyses are sensitive to the moduli of the elements of the PMNS matrix $|U_{\alpha i}|$, and a linear combination of the complex phases. To capture the unitary constraints, analysis tools use the standard parameters $(\theta_{12}, \theta_{23},\theta_{13})$ and $\delta_{CP}$ instead.   This presents a problem when choosing Bayesian priors: since the parameters are not physically motivated, there is no obvious correct choice for the prior distributions on the angles themselves, and it is precisely when making choices of priors over functions on the angles that choosing a parameterisation becomes important. Historically, T2K's Bayesian framework used priors uniform on convenient expressions of the form $\sin^2{\theta_{i3}}$~\cite{freund}, which are proportional or closely related to leading order contributions of the angles to the oscillation probability\footnote{This is not the only choice of trigonometric function on the angles: other Bayesian oscillation fitters, such as NO$\nu$A's, use priors defined on the square sines of the double angles\cite{aria}}. 
Since long-baseline experiments lack $\theta_{12}$ sensitivity but require a constraint on it to make precise measurements of CP-violation~\cite{DentonTheta12}\cite{LopezMorenoTheta12}, analysers are forced to impose a constraint on this angle. T2K uses the constraint from the global fit to solar and reactor measurements quoted by the PDG~\cite{pdgLatest} as its $\sin^2\theta_{12}$ prior. 

Of additional note is the Jarlskog invariant, which can be written, $J_{CP}=s_{12}c_{12}s_{23}c_{23}s_{13}c_{13}^2\sin\delta_{CP}$~\cite{Jarlskog}, in the convention where $\theta_{13}$ is always the intermediate angle. It is of particular interest to experiments, as the value of $J_{CP}$ governs CP violation in the lepton sector. $J_{CP}$ is also useful when validating changes of parameterisation: its prior remains invariant for different  Tait-Bryan parameterisations, as long as the priors on the mixing angles take the same form.

\begin{figure}
\begin{tikzpicture}[scale=0.7]
		\node [] (0) at (-4, 4) {};
		\node [] (1) at (-4, -4) {};
		\node [] (2) at (4, 4) {};
		\node [] (3) at (4, -4) {};
		\node [style=Black circle] (28) at (0, 2.5) {};
		\node [style=Black circle] (30) at (2.5, 0) {};
		\node [style=Black circle] (31) at (0, 0) {};
		\node [style=Black circle] (32) at (-2.5, 0) {};
		\node [style=Black circle] (33) at (-2.5, -2.5) {};
		\node [style=Black circle] (34) at (2.5, -2.5) {};
		\node [style=Black circle] (35) at (0, -2.5) {};
		\node [style=Black circle] (36) at (-2.5, 2.5) {};
		\node [] (37) at (3.5, 0) {};
		\node [] (38) at (3.5, -2.5) {};
		\node [] (39) at (-3.5, -2.5) {};
		\node [] (40) at (-3.5, 0) {};
		\node [] (41) at (-3.5, 2.5) {};
		\node [] (42) at (-2.5, 3.5) {};
		\node [] (43) at (0, 3.5) {};
		\node [] (44) at (0, -3.5) {};
		\node [] (45) at (-2.5, -3.5) {};
		\node [] (46) at (2.5, -3.5) {};
		\node [] (47) at (2.5, 3.5) {};
		\node [] (48) at (3.5, 2.5) {};
		\node [] (49) at (-2.5, 4.5) {};
		\node [] (50) at (0, -4.5) {};
		\node [] (51) at (-2.5, -4.5) {};
		\node [] (54) at (4.5, 0) {};
		\node [] (55) at (4.5, -2.5) {};
		\node [] (58) at (0.5, -1.25) {};
		\node [] (59) at (-1.25, 0.5) {};
		\node [] (60) at (-1.25, -3) {};
		\node [] (61) at (-3, -1.25) {};
		\node [] (62) at (-1.25, 0.5) {};
		\node [] (63) at (-2, 0.75) {};
		\node [] (64) at (-0.5, 0.75) {};
		\node [] (65) at (-2, -3.25) {};
		\node [] (66) at (-0.5, -3.25) {};
		\node [] (67) at (0.75, -0.5) {};
		\node [] (68) at (0.75, -2) {};
		\node [] (69) at (-3.25, -2) {};
		\node [] (70) at (-3.25, -0.5) {};
		\node [style=Green circle] (71) at (2.5, 2.5) {};
		\node [style=Green circle] (72) at (2.5, 2.5) {     };
		\node [style=Black circle] (75) at (2.5, 2.5) {};
		\node [] (76) at (3.25, 3.25) {};
		\node [] (78) at (-2.5, -4.5) {};
		\node [] (79) at (0, -5.5) {};
		\node [] (80) at (3, -5.5) {};
		\node [text=red] (81) at (-1.25, -5) {$\theta_{12}$};
		\node [text=blue] (82) at (5.4, -1.25) {$\theta_{23}$};
		\node [text=green] (83) at (3.25, 3.25) {$\theta_{13}$};
		\node [style=Green circle, minimum size=6mm] (72) at (2.5, 2.5) {     };
		\node [style=Black circle] (28) at (0, 2.5) {};
		\node [style=Black circle] (30) at (2.5, 0) {};
		\node [style=Black circle] (31) at (0, 0) {};
		\node [style=Black circle] (32) at (-2.5, 0) {};
		\node [style=Black circle] (33) at (-2.5, -2.5) {};
		\node [style=Black circle] (34) at (2.5, -2.5) {};
		\node [style=Black circle] (35) at (0, -2.5) {};
		\node [style=Black circle] (36) at (-2.5, 2.5) {};
		\node [style=Green circle] (71) at (2.5, 2.5) {};
		\node [style=Black circle] (75) at (2.5, 2.5) {};
		\node [text=white, font={\bfseries}] (84) at (-1.25, -1.25) {$\delta_{CP}$};
		\node [text=blue] (85) at (-5.6, -2.5) {$\nu_c$};
		\node [text=blue] (86) at (-5.6, 0) {$\nu_b$};
		\node [text=green] (87) at (-5.6, 2.5) {$\nu_a$};
		\node [text=red] (88) at (-2.6, 5.5) {$\nu_x$};
		\node [text=red] (89) at (-0.1, 5.5) {$\nu_y$};
		\node [text=green] (90) at (2.4, 5.5) {$\nu_z$};
		\node [] (91) at (-2.5, 5) {};
		\node [] (92) at (0, 5) {};
		\node [] (93) at (2.5, 5) {};
		\node [] (94) at (0, 4.5) {};
		\node [] (95) at (2.5, 4.5) {};
		\node [] (96) at (-4.5, 2.5) {};
		\node [] (97) at (-4.5, 0) {};
		\node [] (98) at (-4.5, -2.5) {};
		\node [] (99) at (-5, 2.5) {};
		\node [] (100) at (-5, 0) {};
		\node [] (101) at (-5, -2.5) {};
		\draw (2.center) to (3.center);
		\draw (3.center) to (1.center);
		\draw (1.center) to (0.center);
		\draw (0.center) to (2.center);
		\draw [style=Red edge] (36) to (32);
		\draw [style=Red edge] (32) to (33);
		\draw [style=Red edge] (28) to (31);
		\draw [style=Red edge] (31) to (35);
		\draw [style=Blue line] (34) to (33);
		\draw [style=Blue line] (32) to (30);
		\draw [style=Blue line] (32) to (40.center);
		\draw [style=Blue line] (33) to (39.center);
		\draw [style=Blue line] (34) to (38.center);
		\draw [style=Blue line] (30) to (37.center);
		\draw [style=Red edge] (35) to (44.center);
		\draw [style=Red edge] (33) to (45.center);
		\draw [style=Red edge] (36) to (42.center);
		\draw [style=Red edge] (28) to (43.center);
		\draw [style=red arrow] (51.center) to (50.center);
		\draw [style=Blue arrow] (54.center) to (55.center);
		\draw [style=Orange line] (60.center)
			 to [bend right=45, looseness=2.00] (58.center)
			 to [bend right=45, looseness=2.00] (62.center)
			 to [bend right=45, looseness=2.00] (61.center)
			 to [bend right=45, looseness=2.00] cycle;
		\draw [style=Red edge] (42.center) to (45.center);
		\draw [style=Red edge] (44.center) to (43.center);
		\draw [style=Blue line] (38.center) to (39.center);
		\draw [style=Blue line] (40.center) to (37.center);
		\draw [style=new edge style 0, in=180, out=0] (63.center) to (64.center);
		\draw [style=new edge style 0] (66.center) to (65.center);
		\draw [style=new edge style 0] (70.center) to (69.center);
		\draw [style=new edge style 0] (68.center) to (67.center);
		\draw [style=axis2] (91.center) to (49.center);
		\draw [style=axis2] (92.center) to (94.center);
		\draw [style=axis2] (93.center) to (95.center);
		\draw [style=axis2] (99.center) to (96.center);
		\draw [style=axis2] (100.center) to (97.center);
		\draw [style=axis2] (101.center) to (98.center);
 		\node [style=Green circle, minimum size=6mm] (72) at (2.5, 2.5) {     };
		\node [style=Black circle] (28) at (0, 2.5) {};
		\node [style=Black circle] (30) at (2.5, 0) {};
		\node [style=Black circle] (31) at (0, 0) {};
		\node [style=Black circle] (32) at (-2.5, 0) {};
		\node [style=Black circle] (33) at (-2.5, -2.5) {};
		\node [style=Black circle] (34) at (2.5, -2.5) {};
		\node [style=Black circle] (35) at (0, -2.5) {};
		\node [style=Black circle] (36) at (-2.5, 2.5) {};

		\node [style=Green circle] (71) at (2.5, 2.5) {};

		\node [style=Black circle] (75) at (2.5, 2.5) {};
		\node [text=white, font=\bfseries] (84) at (-1.25,-1.25) {$\delta_{CP}$};
\end{tikzpicture}
\caption{Schematic of the mixing matrix generated by expanding $R_{23}\Gamma_\delta R_{13} \Gamma_\delta^\dagger R_{12}$ between the $(\nu_x,\nu_y,\nu_z)$ and $(\nu_a,\nu_b,\nu_c)$ bases. $\theta_{12}$ is a rotation that mixes the $\nu_x$ and $\nu_y$ columns (red), and $\theta_{23}$ is a rotation that mixes the $\nu_b$ and $\nu_c$ rows (blue). $\theta_{13}$ measures the magnitude of the only element untouched by the other angles (green), and $\delta_{CP}$ governs the diagonality/anti-diagonality of the cofactor matrix to the $\theta_{13}$ element (yellow). \label{fig:PMNSschematic}}
\end{figure}
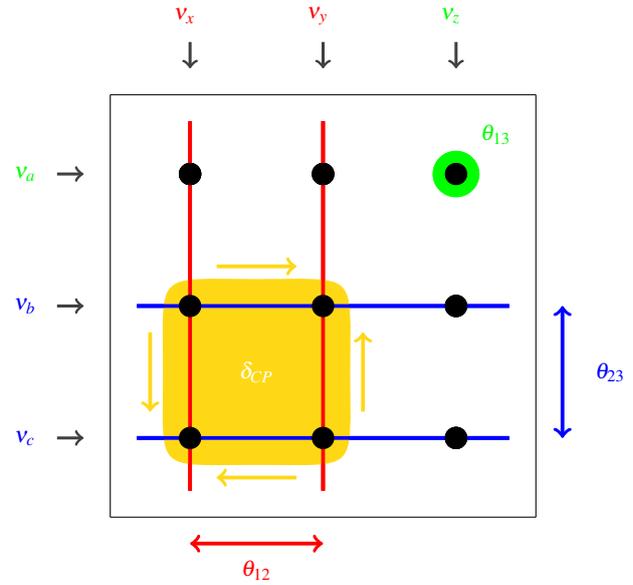

Using a generic Tait-Bryan parameterisation as defined in expression \ref{eqn:genericPars}, the $R_{23}$ matrix is a rotation of a plane defined by two states in some basis $\nu_a,\nu_b,\nu_c$ (in the usual parameterisation, the flavour basis), and $R_{12}$ is a rotation of a plane defined by two states in some other basis $\nu_x,\nu_y,\nu_z$ (in the usual parameterisation, the mass basis). Assuming (without loss of generality), that these are the 2nd and 3rd of the $\{a,b,c\}$ basis, and 1st and 2nd states of the $\{x,y,z\}$ basis, the relation between the mixing parameters and elements of the mixing matrix is described in Figure \ref{fig:PMNSschematic}. A key takeaway from this construction is that the $\theta_{13}$ angle has a different relation to the moduli of the mixing matrix than $\theta_{12}$ and $\theta_{23}$. 

\begin{figure}
    \centering
    \includegraphics[width=0.47\textwidth]{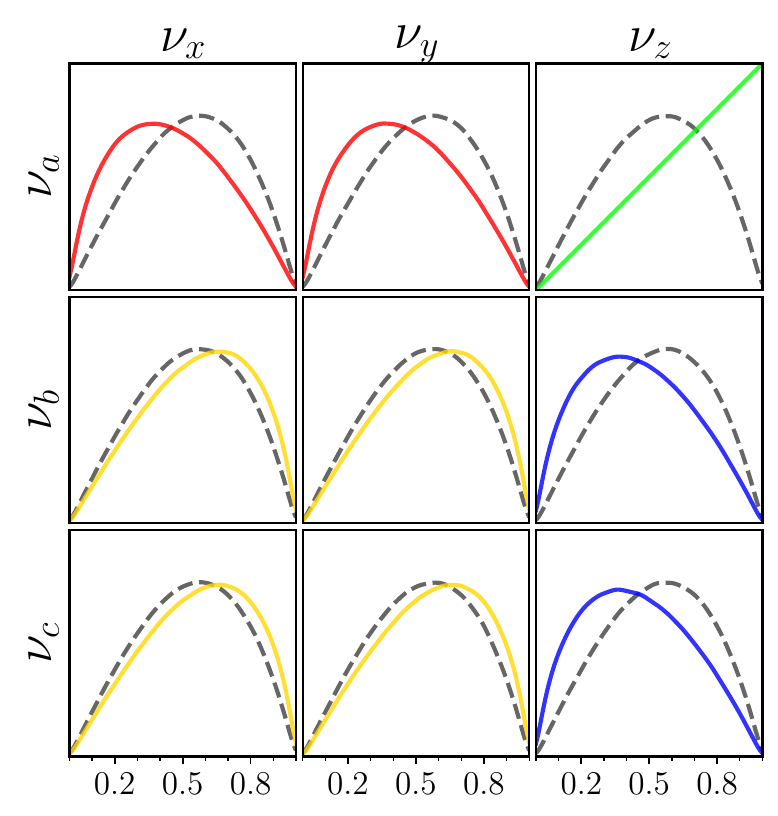}
    \caption{Prior distributions on $|U_{R}|_{ij}$ resulting from uniform priors on $\sin^2\theta_{ij}^X$ and $\delta_{CP}^X$  (solid lines) and expected distribution of $|U_{R}|_{ij}$ for random U(3) matrices as given by the Haar measure (dashed lines). The upper-right distribution (governed by $\theta_{13}^X$) breaks the 9-fold symmetry of the Haar-induced prior. The colours mirror the convention used in figure \ref{fig:PMNSschematic} to highlight the effect of the parameterisation.}
    \label{fig:uniformPriors}
\end{figure}

Now, consider priors uniform over the angles or, as is common in long-baseline analysis, over the square of the sines. These distributions induce uneven priors on the elements of the mixing matrix because not all elements are related to the mixing parameters in the same way (see Figure \ref{fig:uniformPriors}). 

Such priors (uniform in the $\sin^2\theta_{ij}$ and $\delta_{CP}$ of the host parameterisation) will be referred to as Tait-Bryan priors, owning to the name of the parameterisation they are constructed on. These are interesting because they enable us to privilege flavour and mass symmetries, but to gauge their bias it is useful to compare them to a more general prior that lacks structural preferences. One good choice is the uniform prior in the Haar measure~\cite{HaarAnarchy}\cite{HaarPMNS}, which uses the topological group structure of U(3) to create an invariant volume element. The Haar prior corresponds to the distribution we should expect random unitary $3\times 3$ matrices to follow, and is the natural choice if we assume the PMNS has no structural preferences. The hypothesis described by the Haar prior is commonly referred to as flavour anarchy~\cite{anarchy}, because it represents the antithesis to flavour hierarchies. 

\begin{figure}
    \centering
    \includegraphics[width=.47\textwidth]{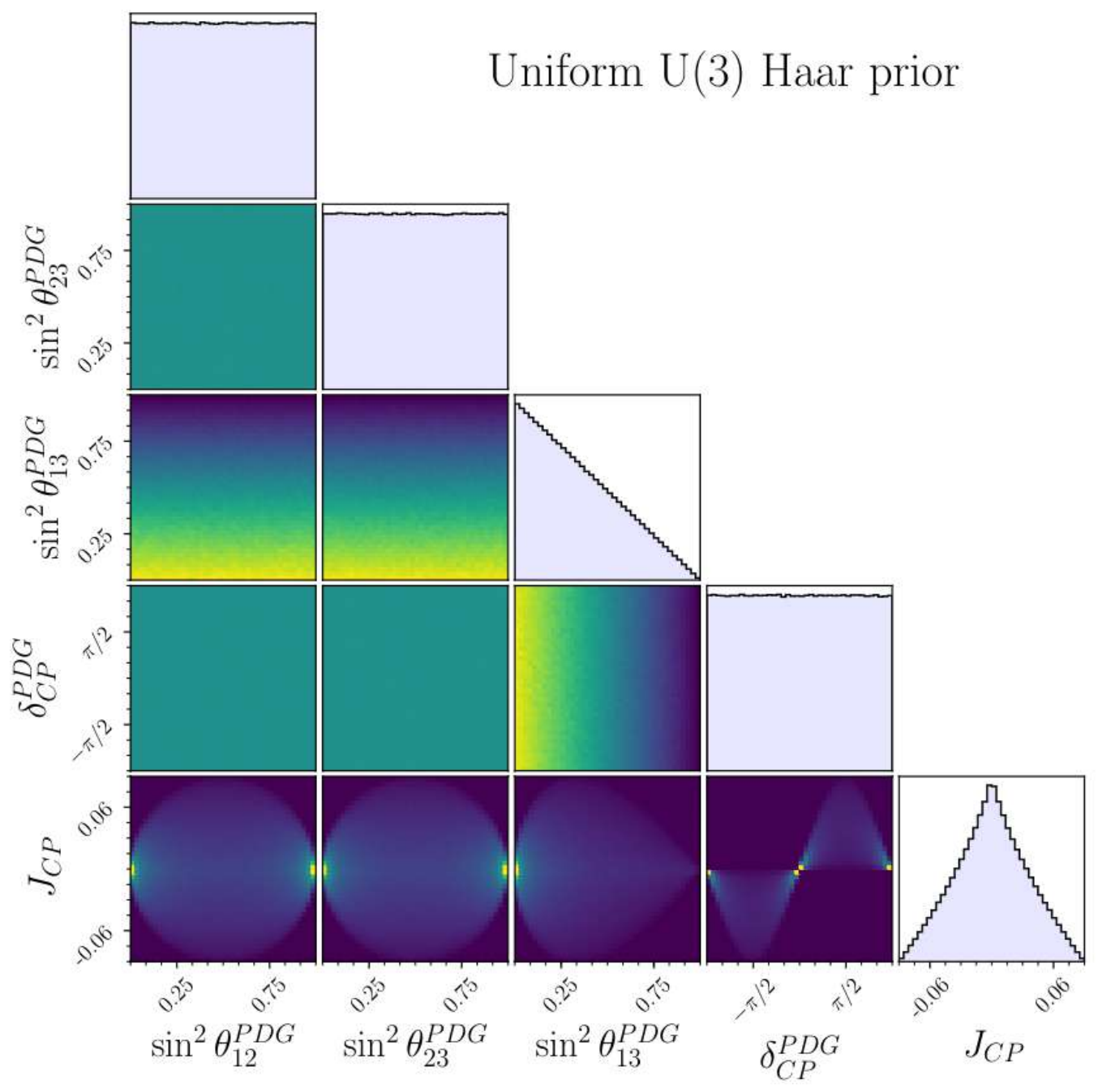}
    \caption{1D and 2D marginalised priors on the standard parameters induced by the Haar measure of the U(3) matrix space. $J_{CP}$ is the Jarlskog invariant and is not a free parameter.}
    \label{fig:HaarPrior}
\end{figure}

The Haar prior can be written in terms of the parameters of a Tait-Bryan parameterisation as uniform in the squared sines of the rotational angles $\sin^2\theta_{12}, \sin^2\theta_{23}$, uniform in the quartic cosine of the third angle $\cos^4\theta_{13}$, and uniform in $\delta_{CP}$; its 1D and 2D projections onto the standard parameters are shown in Figure \ref{fig:HaarPrior}, but one should keep in mind that these are correlated across the 4D parameter space. When written in terms of the elements of the mixing matrix, they all follow the same distribution $\pi_\text{Haar}(U_{ij})= 4|U_{ij}|(1-|U_{ij}|^2)$.

When compared to the flavour anarchic Haar prior, uniform priors in Tait-Bryan parameters tend to highlight symmetries between the rows and columns containing the states in the rotation planes of $\theta_{12}$ and $\theta_{23}$. In particular, the canonical parameterisation induces a prior skewed towards $\nu_\mu/\nu_\tau$ and $\nu_1/\nu_2$ symmetries. This is made apparent in Figure \ref{fig:uniformPriors}, where the shape of the priors corresponds with the parameterisation-coded colouring from Figure \ref{fig:PMNSschematic}.

When attempting a Bayesian fit in long-baseline oscillation analysis, if the analyser strongly believes in a structureless mixing matrix, the correct prior is the Haar prior above. While there is no reason to believe in symmetries between random non-eigenstates, there is theoretical interest in exact and broken symmetries between mass states and flavour states~\cite{flavourSymReview}\cite{TriMaximal}\cite{TriBimaximal}. We can find Tait-Bryan parameterisations whose uniform priors privilege each choice of flavour and mass symmetry by setting rotation planes that contain the desired states. This leads to 9 such parameterisations, one of which is the canonical scheme. Up to re-labelling of the angles and changing the sign of the complex phase, these 9 parameterisations can be arrived at by changing the combination and ordering of rotation matrices in equation \ref{eqn:PDG} (see appendix \ref{sec:fullPMNS}). This method was used in \cite{Denton} to arrive at the complete matrix expressions, which have been reproduced in \ref{sec:fullPMNS}. Here we study the robustness of T2K's latest results~\cite{oa2023} under priors derived from these 8 additional parameterisations.

\section{MCMC fits in alternate parameterisations}
\label{sec:MCMC}
The reanalysis of T2K's results is performed by weighting steps from a Markov chain Monte Carlo (MCMC) analysis of T2K data, as rerunning the analysis is computationally expensive. To do this, the ratio between the prior used in the original analysis and the new priors is calculated and the posterior distributions are reweighted accordingly. While there is an analytic bijective map between the parameterisations, propagating the alternate prior distributions onto the standard parameters analytically is a challenging and time-consuming task. Instead, the weights are approximated numerically on a grid. This approximation is performed through the binned distribution in the original space of a large (10$^{11}$ draws) uniformly distributed sample drawn from the alternate parameterisation space.

\begin{figure}
    \centering
    \includegraphics[width=0.47\textwidth]{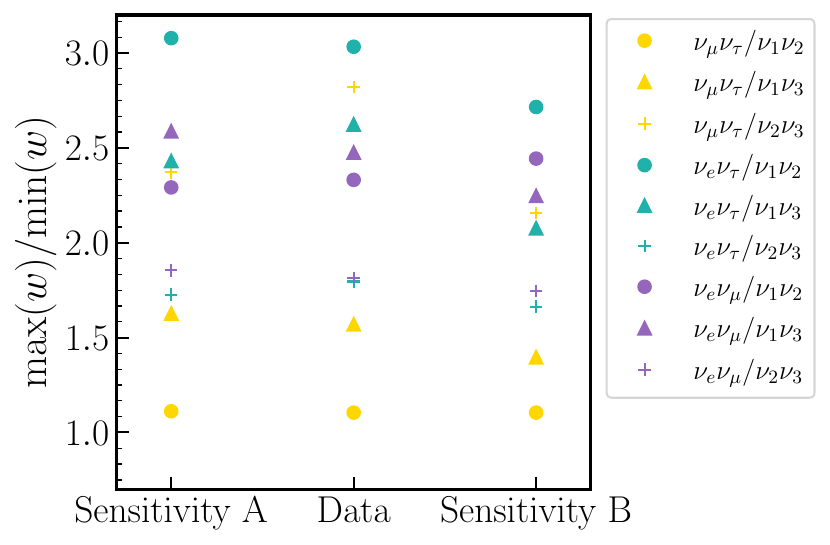}
    \caption{Ratio of the largest over smallest weight applied to the MCMC chains when producing each prior. The ratios are small because the regions of parameter space which would receive the most extreme weights are excluded by the solar and/or T2K constraints. These ratios act as an upper bound for the amplification of the error in the MCMC approximation of the posterior due to the reweighing. The colors indicate the flavour pair of the prior and the shapes indicate the mass pair.}
    \label{fig:weightRatios}
\end{figure}

\begin{figure*}[pt]
    \centering
    \includegraphics[width=0.45\textwidth]{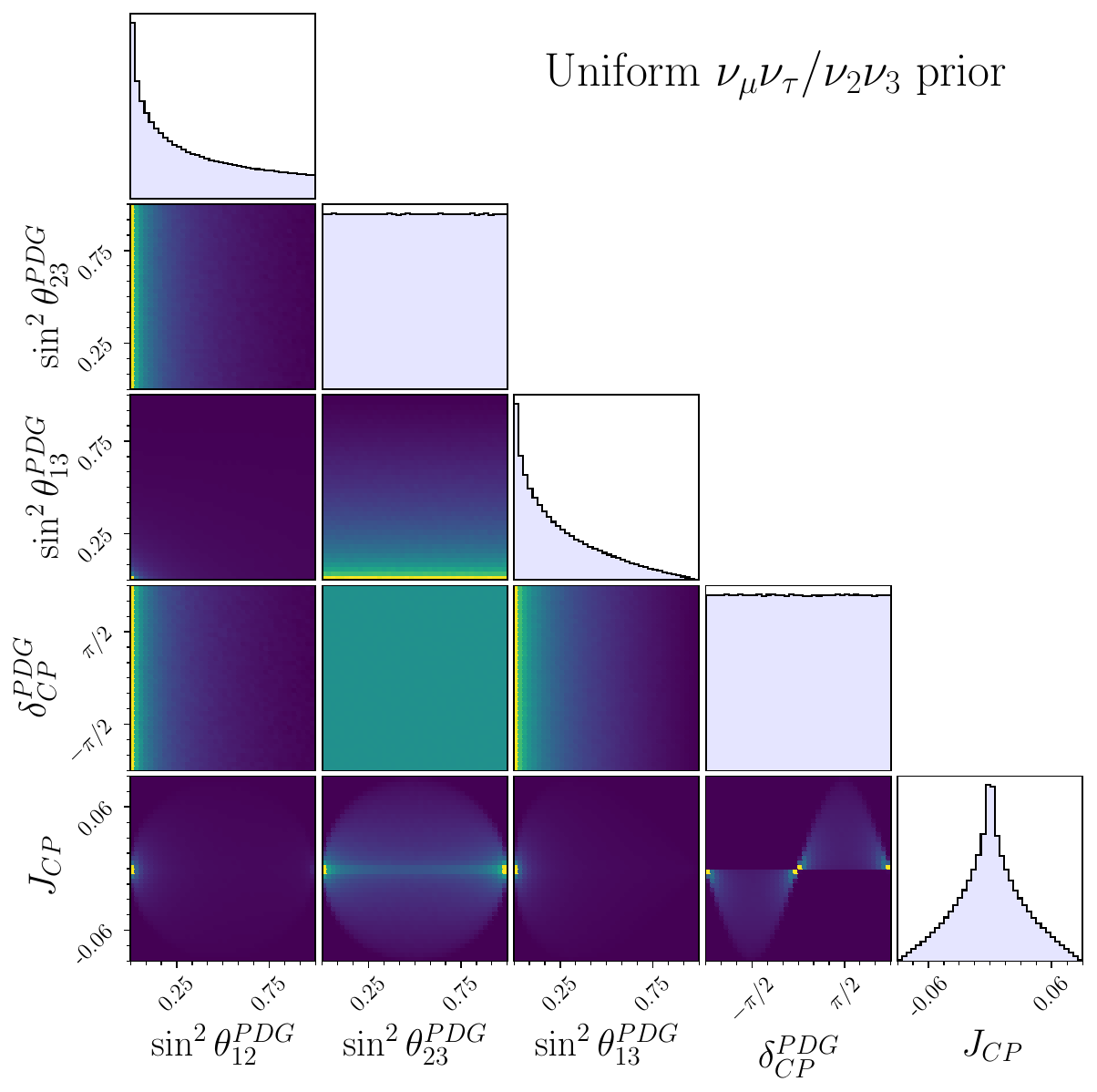}
    \includegraphics[width=0.45\textwidth]{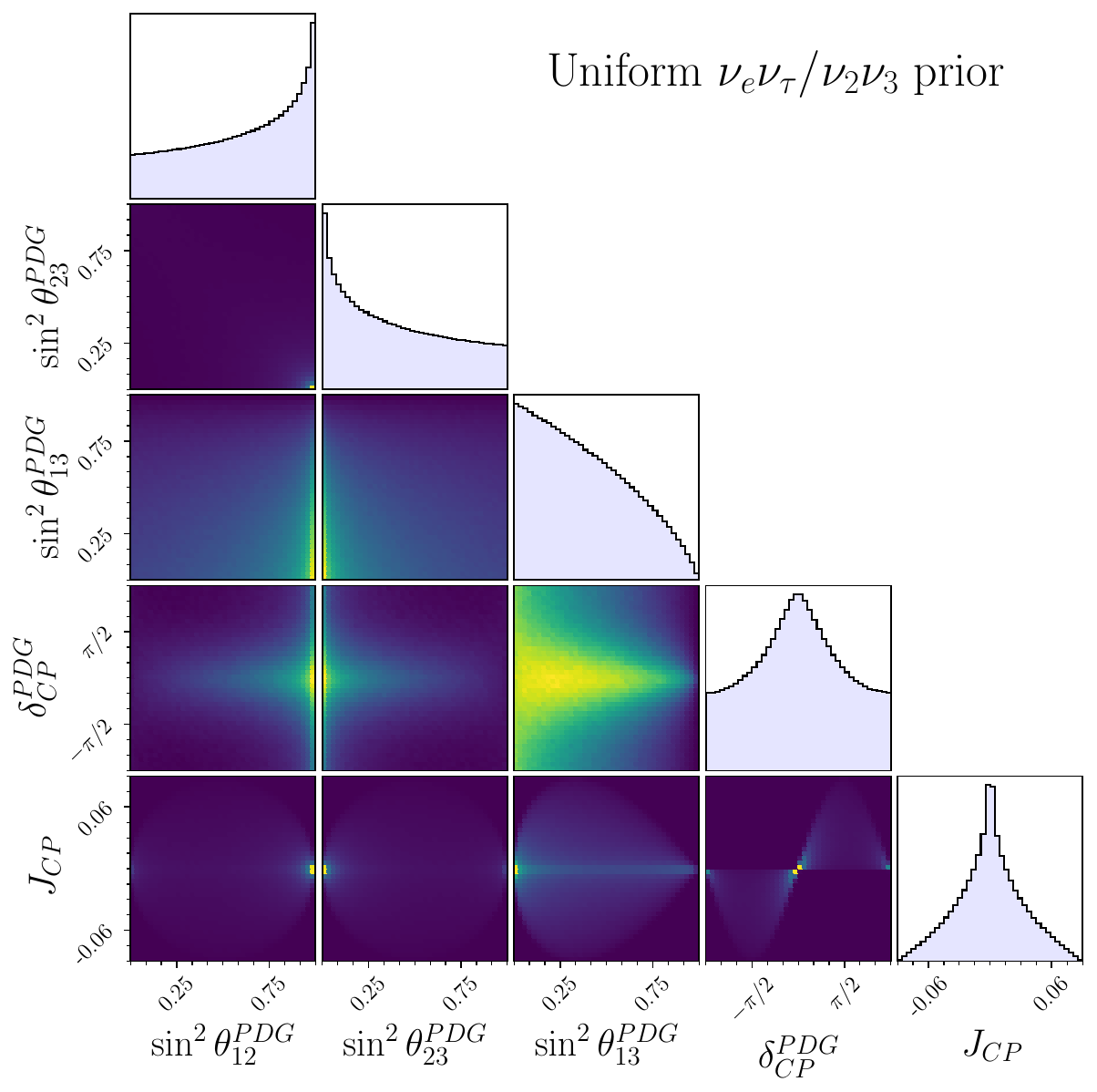}
    
    \includegraphics[width=0.45\textwidth]{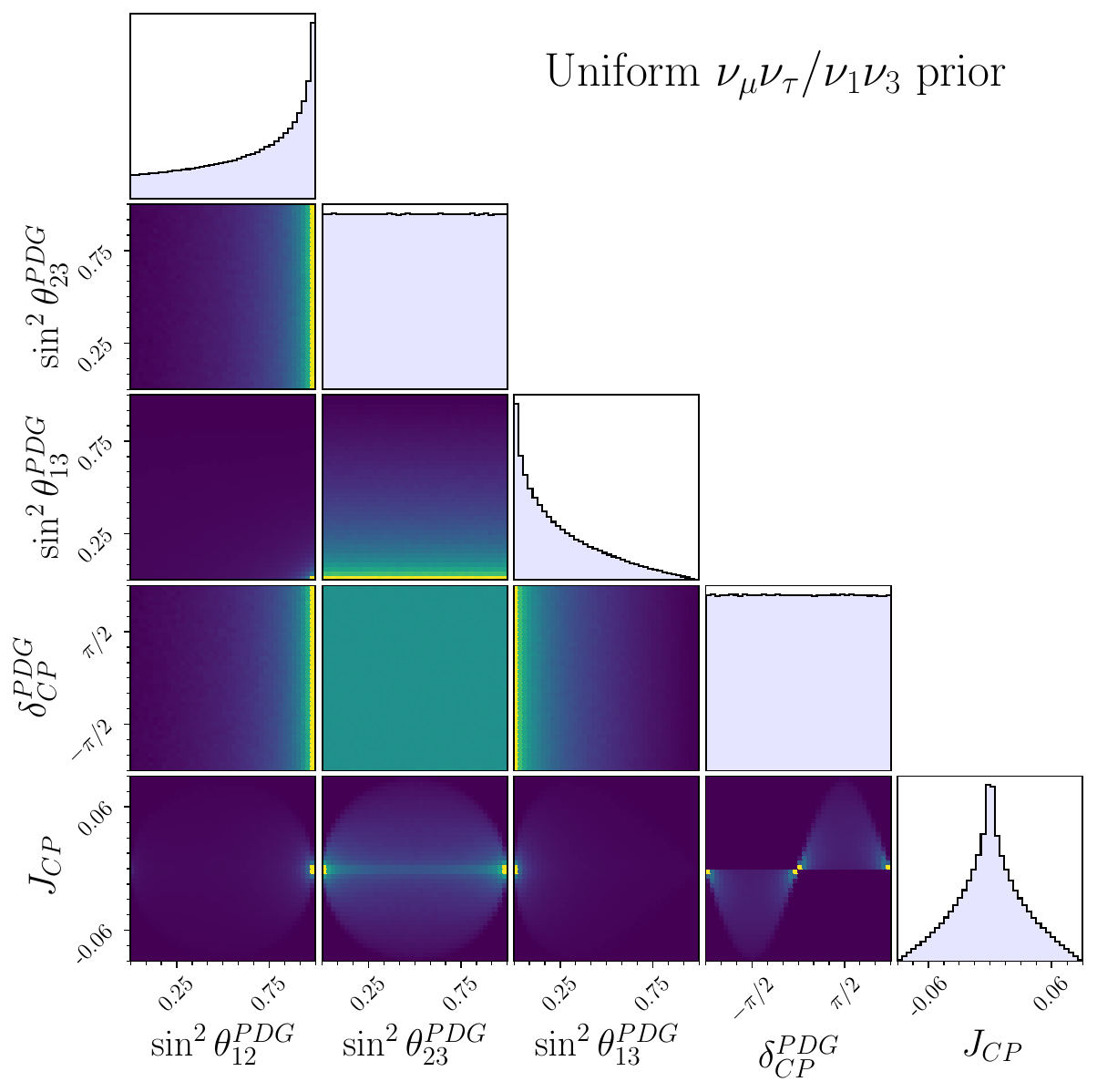}
    \includegraphics[width=0.45\textwidth]{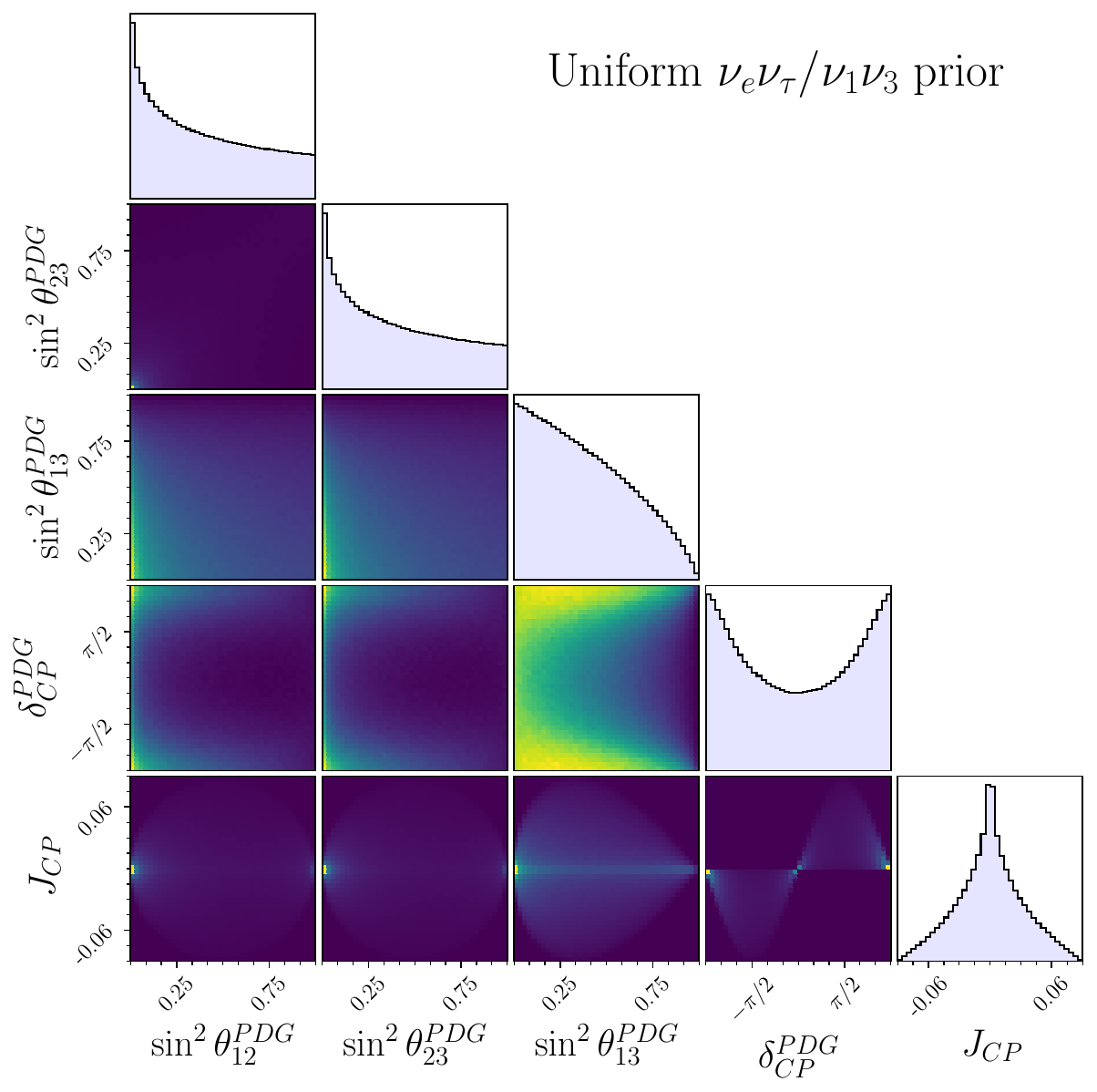}
    
    \includegraphics[width=0.45\textwidth]{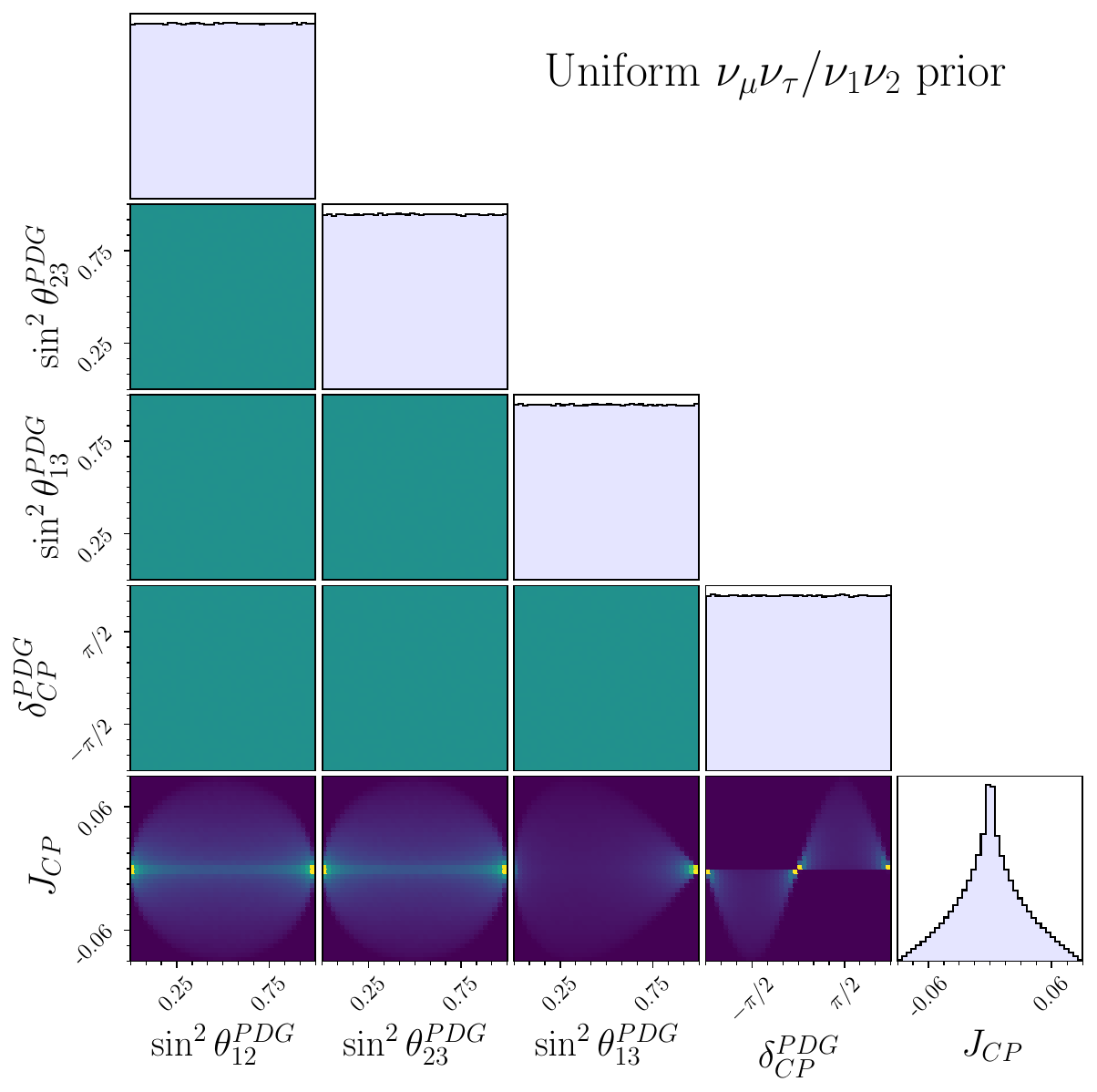} 
    \includegraphics[width=0.45\textwidth]{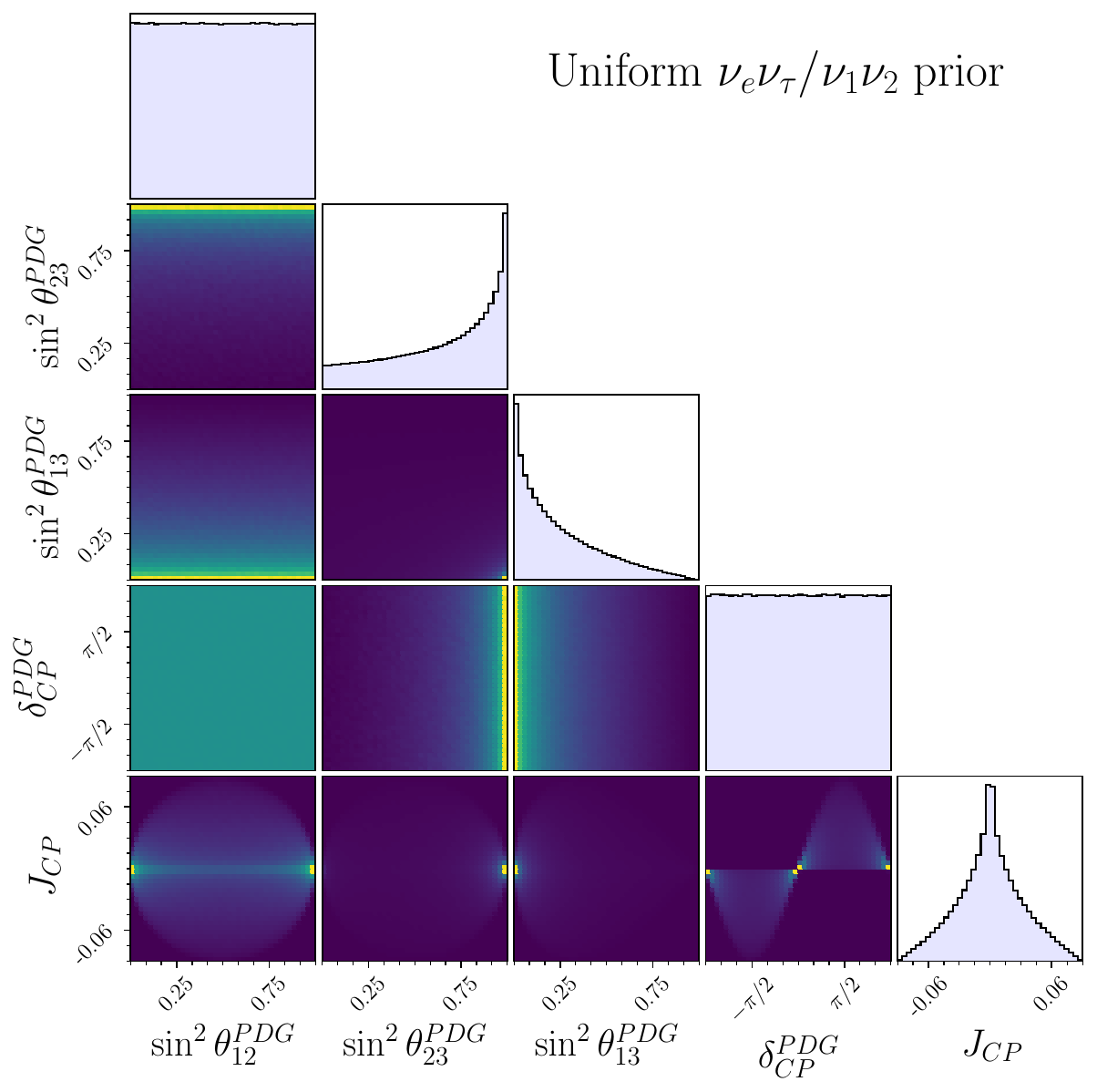}
\end{figure*}
\begin{figure*}[pt]
    \centering
    \includegraphics[width=0.45\textwidth]{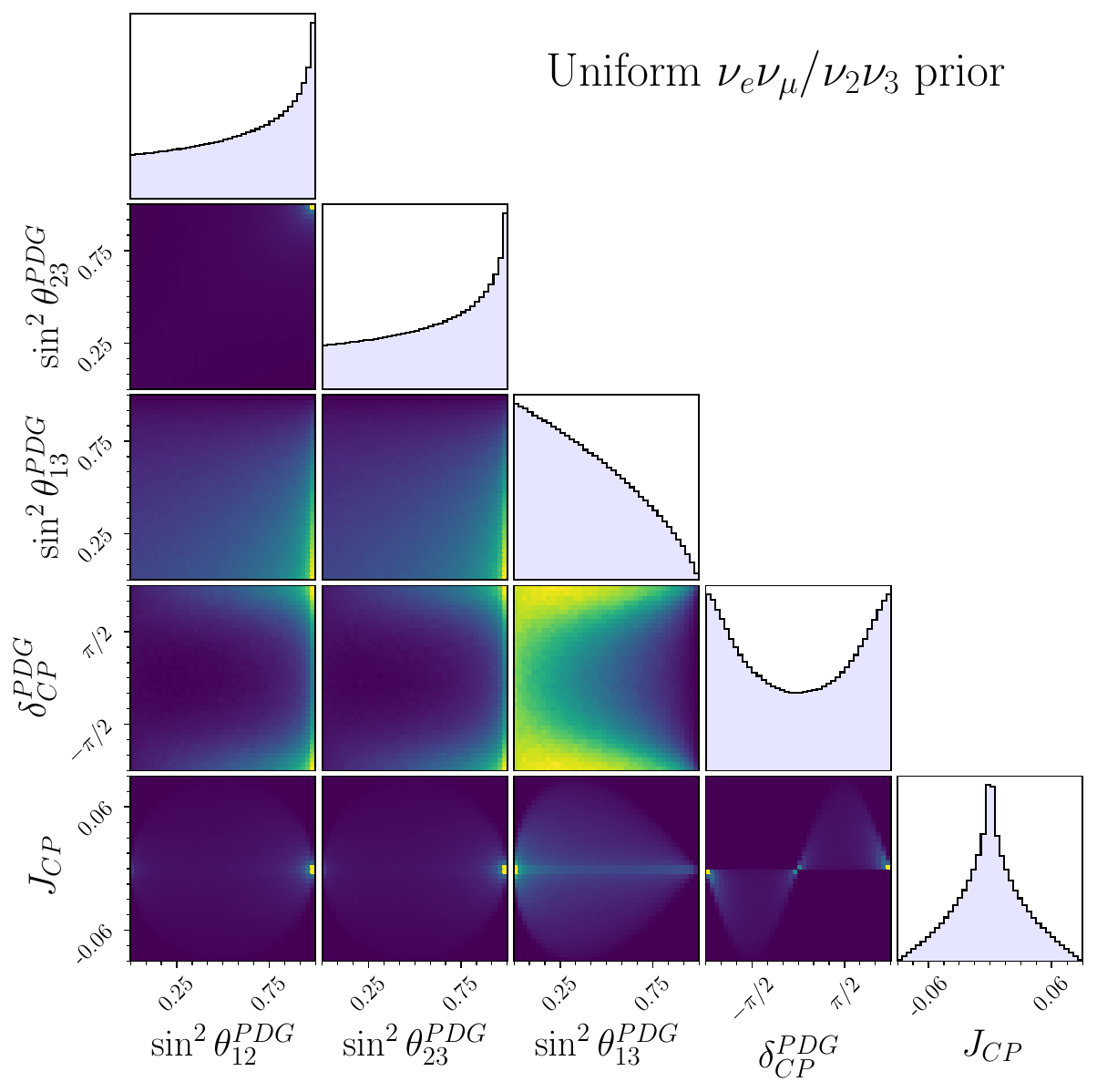}
    \includegraphics[width=0.45\textwidth]{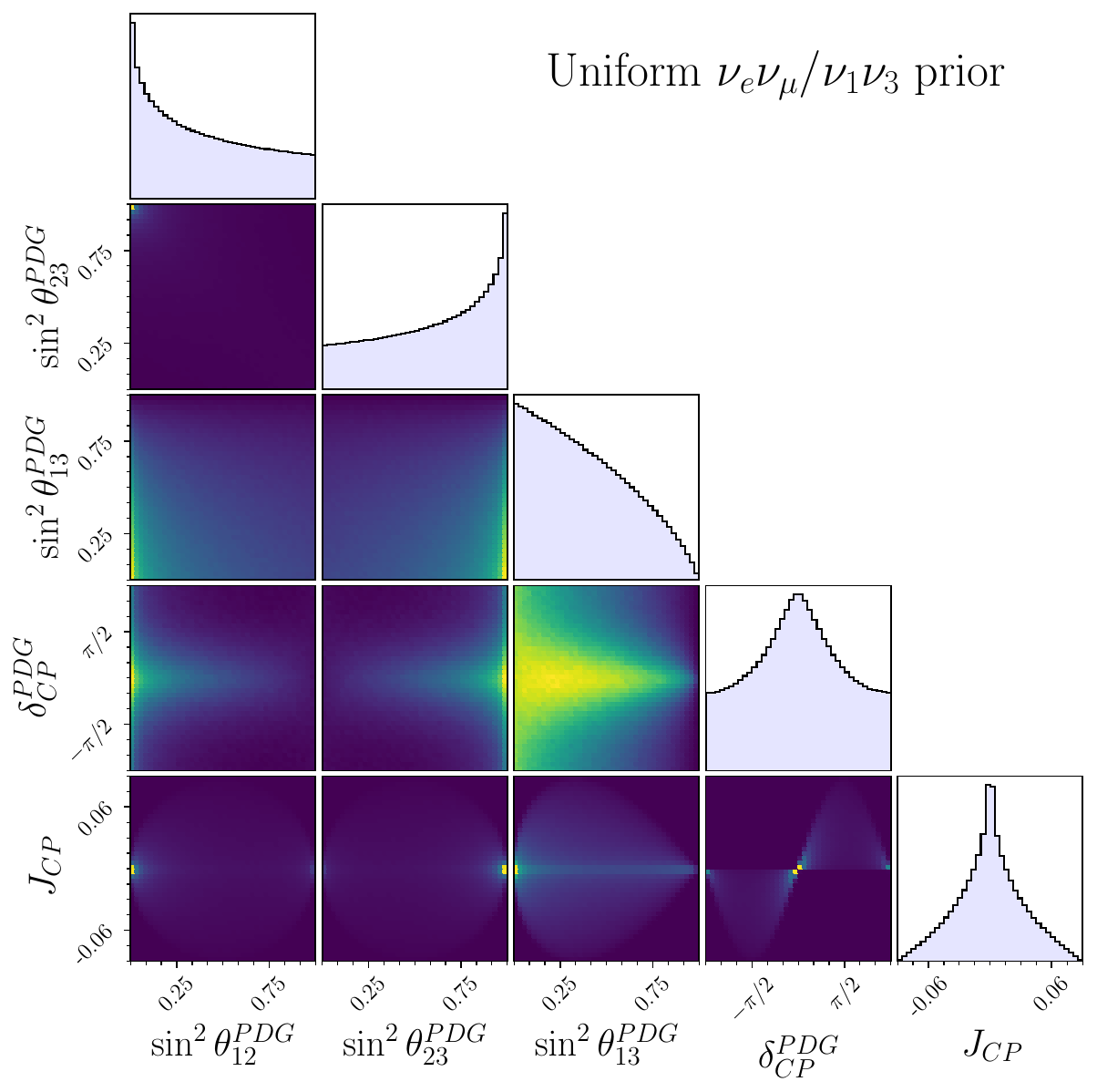}
    \includegraphics[width=0.45\textwidth]{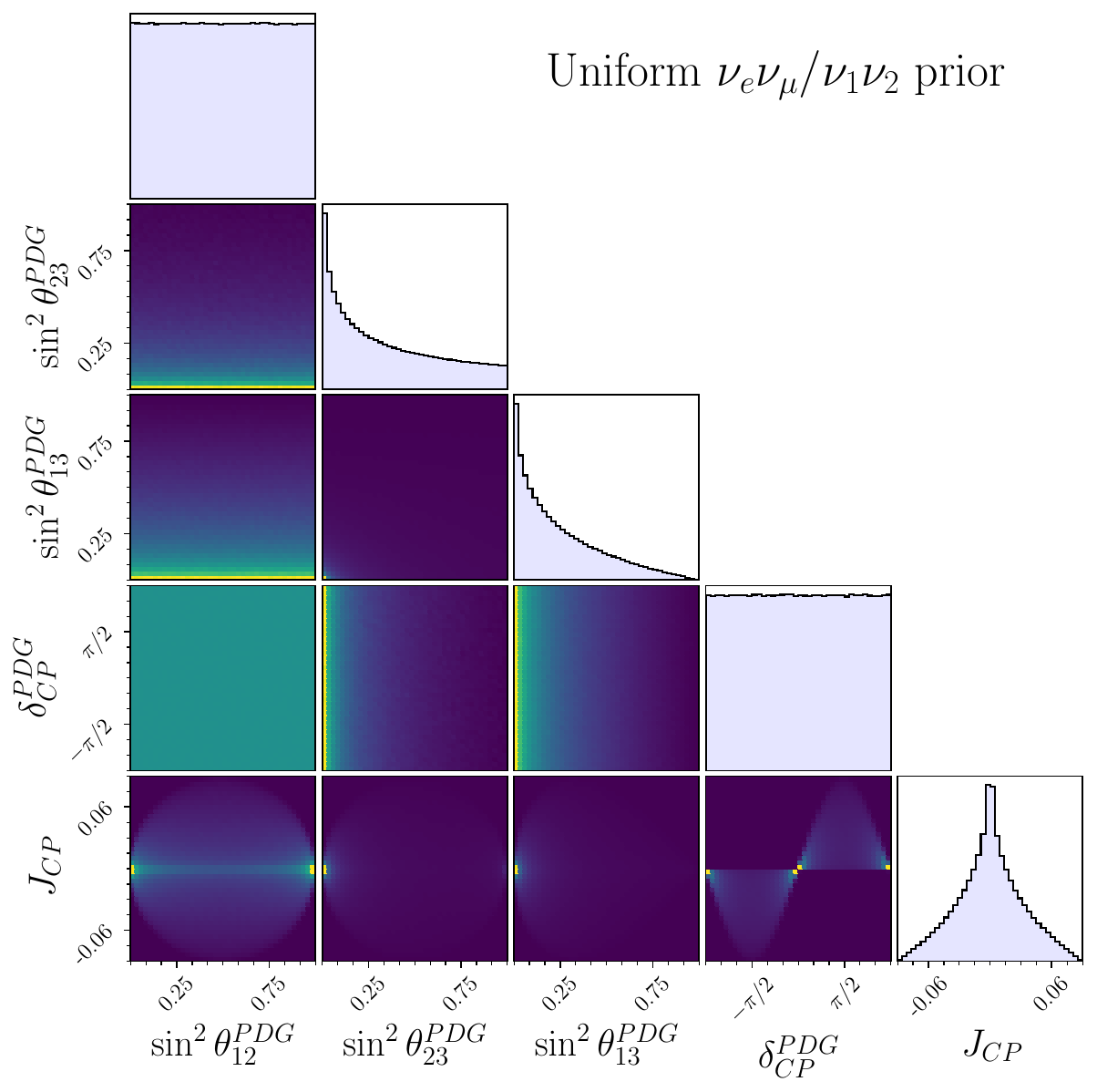}
    \caption{Marginalised 1D and 2D Tait-Bryan priors over the standard parameters. Each prior was generated by drawing uniformly in some alternate Tait-Bryan parameterisation (labelled by the symmetry they privilege). The bottom-left plot on the first page corresponds to the standard parameterisation (here labelled as $\nu_\mu\nu_\tau/\nu_1\nu_2$) and is therefore flat everywhere. The plots are arranged in groups of three (left and right columns on the first page, three plots on the second page) by the flavour symmetry they privilege. The supplementary material displays the nine plots according to the symmetries they privilege, and gives an intuition on how to interpret the priors.}
    \label{fig:priors}
\end{figure*}

This method introduces two sources of uncertainty: statistical uncertainties in the weight approximation, which become negligible in areas of high posterior density\footnote{This is true as long as the assigned weights for steps in the same posterior bin are largely uncorrelated. We ensure this is the case by having many more grid points than steps in our Markov-Chain and studying marginalised 1D and 2D posteriors instead of the complete 4D posterior distribution.}, and amplified uncertainties resulting from giving a large weight to a sparsely populated posterior bin.

Assuming the approximated weights and the posterior bins follow Poisson distributions and the number of steps in any two posterior bins are uncorrelated, the induced uncertainty on the number of steps $n$ in bin $b$ of the reweighted posterior Var$(n(b))$ is
\begin{equation}
    \text{Var}(n(b)) = \frac{N}{W}w(b)p(b) +  \frac{N^2}{W}w(b)p^2(b) + Nw^2(b)p(b)
    \label{eqn:uncertainty}
\end{equation}
where $N$ is the total number of steps in the MCMC chain, $W$ is the number of draws used in the approximation of the weights, $w(b)$ is the true weight of bin $b$ and $p(b)$ is the true value of the unweighted posterior at bin $b$.

The first two terms can be made arbitrarily small by taking a large sample in the calculation of the weights; in this study, their contribution is kept at below 1\% of the original bin uncertainty for the entire $3\sigma$ range. We can find an upper bound for the final term by assuming the largest weights are given to the low posterior density regions and the smallest weights are assigned to the highest posterior density bins. In this scenario, the final term is at most
\begin{equation}
    \sqrt{Nw^2(b)p(b)}\leq \frac{\max(w(b'))}{\min(w(b'))}\sqrt{Np(b)}
    \label{eqn:upperBound}
\end{equation}
that is, the ratio of largest to smallest weight applied to an MCMC chain serves as a (conservative) upper bound for the amplification factor on the variance of the posterior approximation introduced by the new weights. 

Figure \ref{fig:weightRatios} shows the ratio between the largest and smallest applied weights of each parameterisation for two sensitivity analyses and one data chain\footnote{The oscillation parameters used in the sensitivity fits are presented in \ref{sec:fakeResults}}. Taking the largest ratio as an upper limit of the amplification, the new uncertainty is at most 3 times larger than the original; this is satisfactory because the statistical uncertainty within the $3\sigma$ region of our $2\times10^8$ step posterior chains is sub-percent.

\section{Results}
\label{sec:results}
\begin{figure}
    \centering
    \includegraphics[width=.47\textwidth]{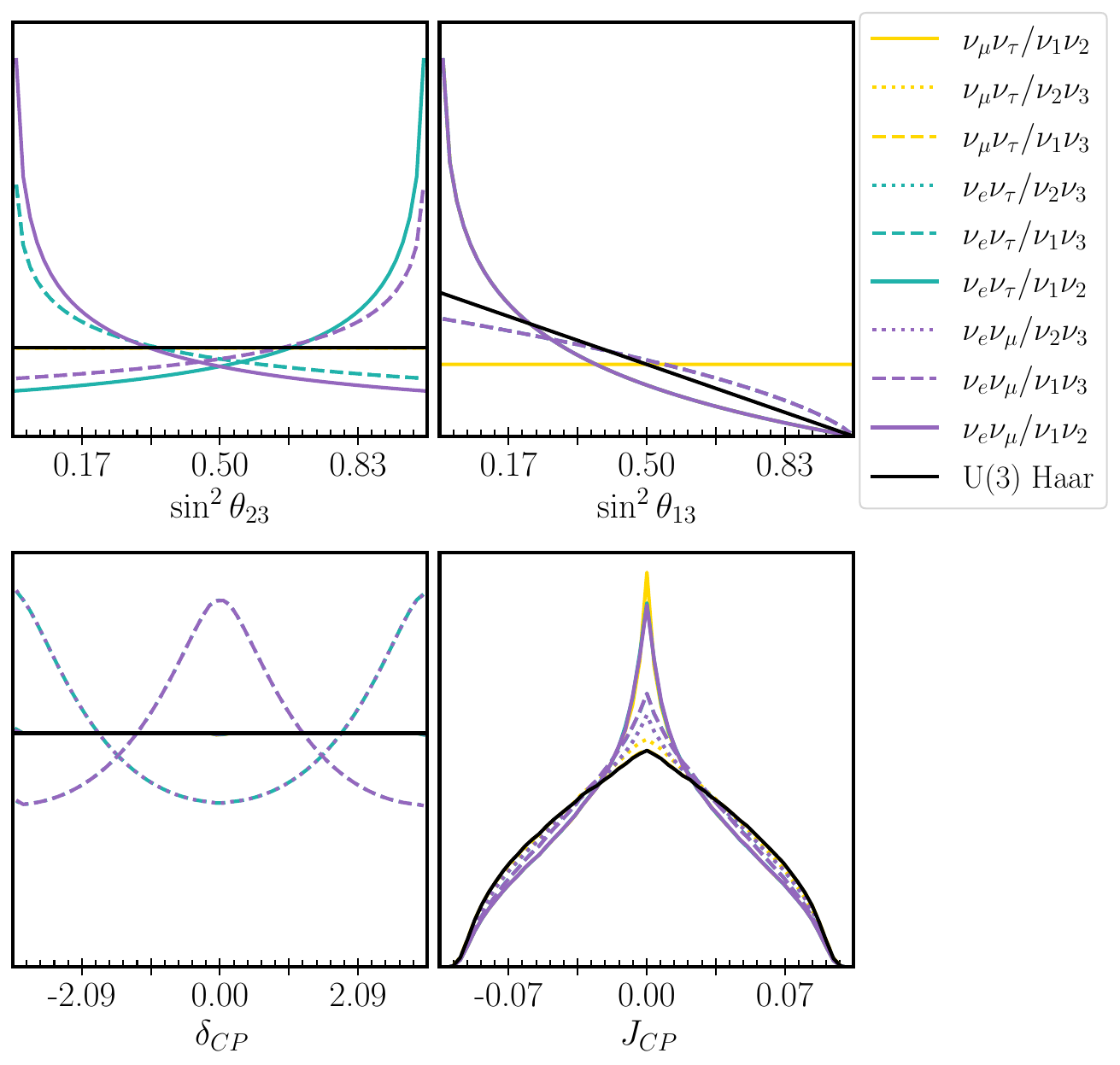}
    \caption{Priors uniform in each of the 9 flavour/mass symmetric Tait-Bryan parameterisations and Haar prior for the relevant oscillation parameters, after imposing a Gaussian prior on $\sin^2\theta_{12}^\text{PDG}$ ($\mu=0.307,\sigma=0.041$) derived from the $\theta_{12}$ constraint reported by the PDG.  The priors are labelled by the symmetries they privilege, and the standard PDG prior is $\nu_\mu\nu_\tau/\nu_1\nu_2$. The colours indicate the flavour pair of the prior, and the line styles indicate the mass pair.\label{fig:1Dpriors}}
\end{figure}

Figure \ref{fig:priors} shows the 1D and 2D marginalised priors derived from uniform priors in the alternate parameterisations on the standard parameters. Some parameterisations share a uniform prior on $\delta_{CP}^\text{PDG}$ while others favour CP conservation by up to $15\%$. This is the consequence of a re-definition of the complex phase happening in those parameterisations where the elements with a complex component swap places with the purely real ones\footnote{This is allowed because oscillations are only sensitive to specific linear combinations of the phases.}. Since $J_{CP}$ takes the same from under all parameterisations, every Tait-Bryan prior must assign to it the same distribution. This is precisely what our computation shows, and serves as a sanity check when changing parameterisations.

We apply the solar constraint by imposing a Gaussian prior on $\sin^2\theta_{12}^\text{PDG}$ taken from the global solar constraint in the PDG report \cite{pdgLatest} (as is usual in T2K analyses). Doing so on top of uniform priors on each parameterisation breaks the invariance and leads to different prior distributions on the amount of CP violation. This is evident in Figure \ref{fig:1Dpriors}, where the alternate parameterisation priors have been applied together with the solar constraint.

In future oscillation analyses, to more accurately capture the solar measurement and remove prior reliance on the smallness of $|U_{e3}|$, it is advisable to consider alternative solar constraints. This could be achieved by using priors derived from KamLAND's reactor measurements~\cite{KamLAND} or by playing the reparameterisation game to express the solar measurement in a Tait-Bryan scheme where the simple element falls in $U_{e2}$.

\begin{figure*}
    \centering
    \includegraphics[width=0.9\textwidth]{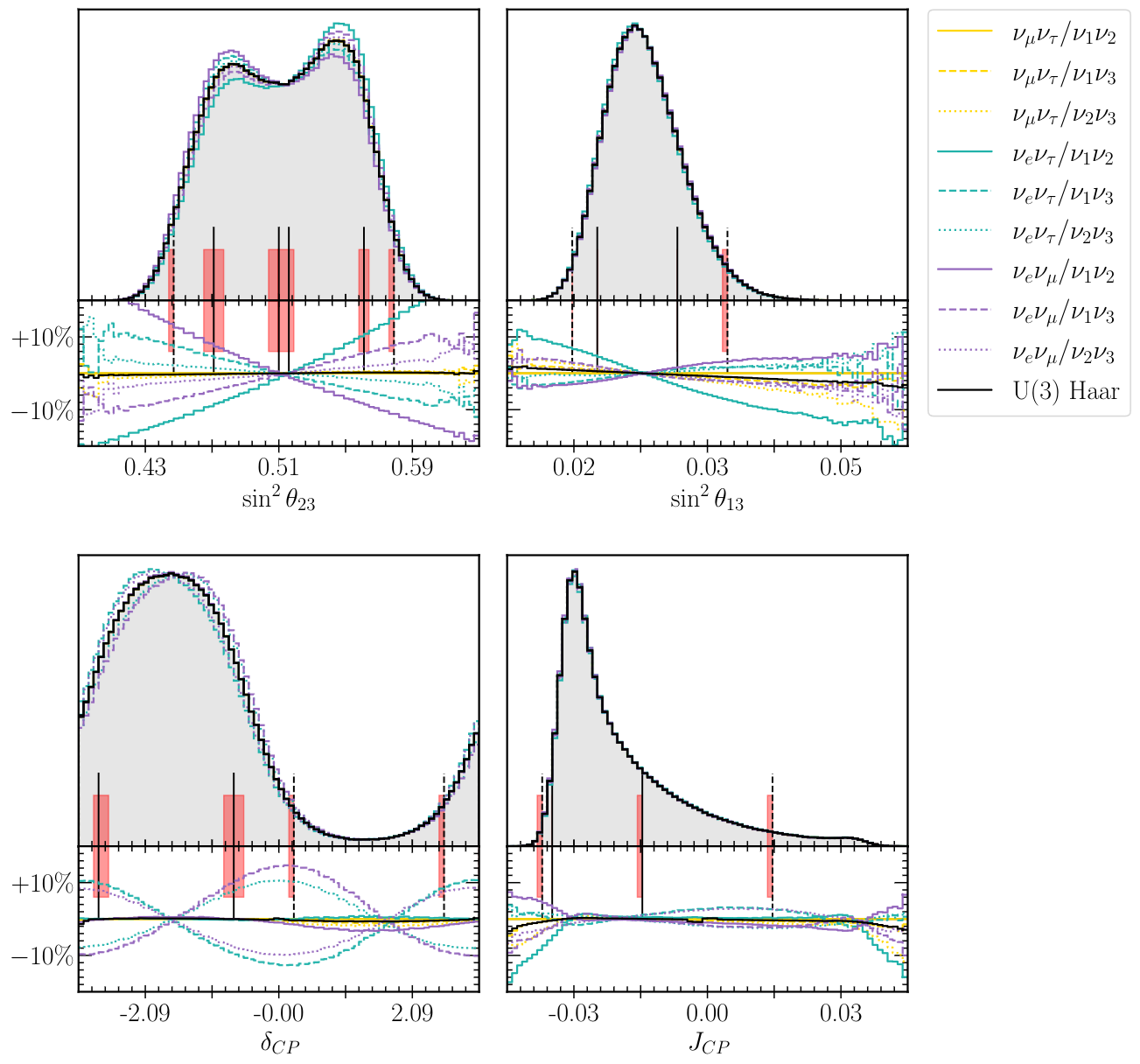}
    \caption{1D marginalised posterior over the standard parameters of T2K's 2022 oscillation analysis, reweighted under the 9 flavour/mass symmetry and Haar priors. The colours indicate the flavour pair of the prior and the line styles indicate the mass pair. The grey area corresponds to the original posterior generated using the PDG prior, and the vertical lines mark the $1\sigma$ (filled) and $2\sigma$ (dashed) credible regions. The red areas include the boundaries of the credible intervals for all the studied priors, and serve as an indication of how much each interval varies. The bottom plot shows the fractional bin change from the standard prior. }
    \label{fig:1Dposterior}
\end{figure*}

\begin{figure*}
    \centering
    \includegraphics[width=.9\textwidth]{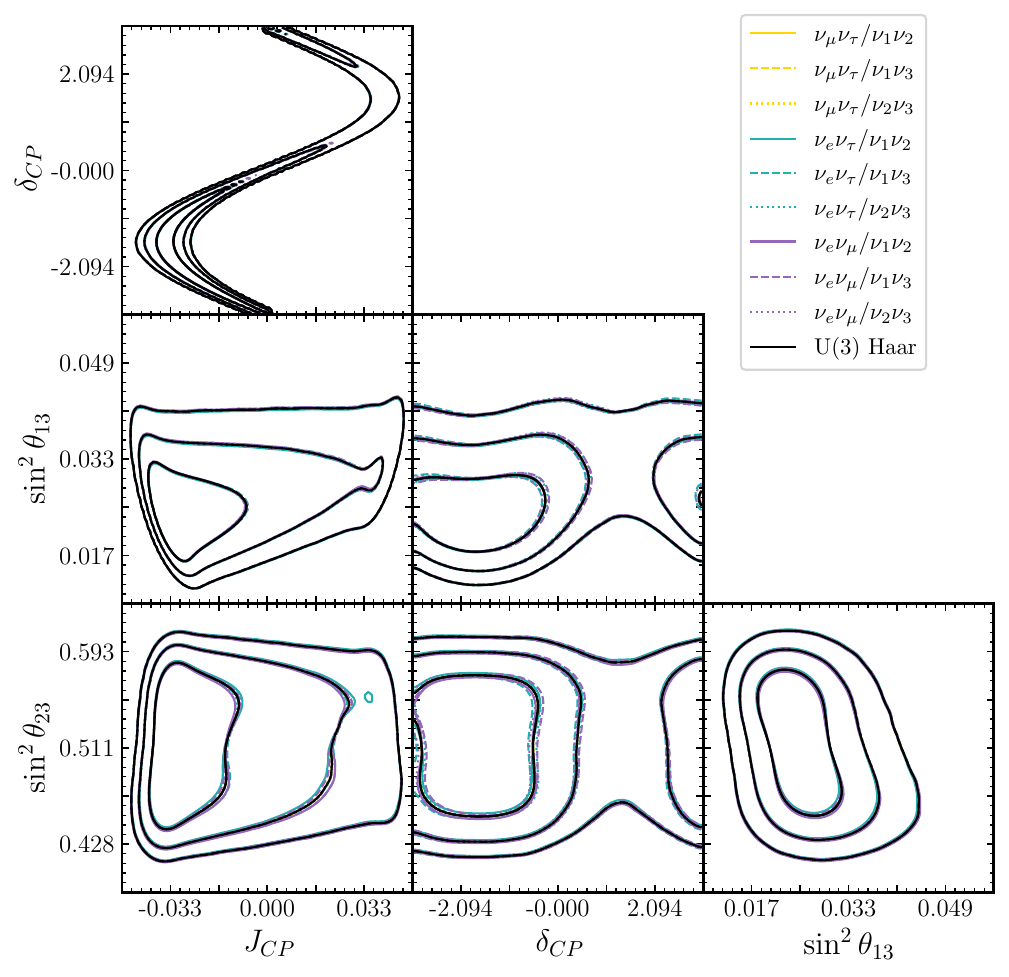}
    \caption{2D marginalised posteriors for T2K's 2022 oscillation analysis, reweighted under the 9 flavour/mass symmetry and Haar priors. The contour lines correspond to the 1$\sigma$, 2$\sigma$, and 3$\sigma$ credible regions.}
    \label{fig:2Dposteriors}
\end{figure*}

Figures \ref{fig:1Dposterior} and \ref{fig:2Dposteriors} show the 1D and 2D marginalised posteriors resulting from applying the Tait-Bryan priors, together with the solar constraint, to T2K's 2022 oscillation analysis, which improves on the analysis presented in~\cite{oa2023}. Although the priors vary significantly (Figure \ref{fig:1Dpriors}), the credible regions show only small variations from the original fit. 

A particularly interesting way to quantify the prior dependence on T2K's physics conclusions is to ask how it affects the credible intervals. Despite the fractional bin-by-bin differences being up to $\approx 10\%$, meaningful variations in the intervals only appear in the $\sin^2\theta_{23}^\text{PDG}$ posteriors. In this latter case, several alternate priors result in an enhancement of the posterior in the lower octant, indicating that the weak upper-octant preference of the original analysis can be interpreted as a result of the choice of prior. 

In terms of CP-violation, while the marginalised posteriors in $\delta_{CP}$ show some variation, the credible intervals over the Jarlskog invariant stay effectively constant. Since it is more closely related to the experimental event rates than the mixing angles~\cite{T2Knature}, it is not surprising to see that the data imposes more robust constraints on $J_{CP}$ than on the individual mixing parameters. This serves as a reminder that the Jarlskog invariant is the true measure of CP-violation and $\delta_{CP}$ constraints do not give the full picture and shows that T2K's evidence for CP violation is robust under these choices of prior. 


\ref{sec:fakeResults} presents the results from running this same analysis on two additional MCMC posteriors which come from sensitivity analyses. The simulated data for the Asimov A MCMC chain were generated using parameter values similar to T2K's best fit, and serve to confirm that these results are not an artefact of some undetected tensions between T2K samples. The Asimov B MCMC chain uses vastly different parameter values (though still consistent with existing data) and serves to verify that the small difference in the posteriors is a consequence of T2K's strong constraining power and not an artefact of the region of parameter space favoured by current data.

\section{Conclusion}
This work discussed the space of Tait-Bryan parameterisations of the lepton mixing matrix and their relation to row-column symmetries. We showed that uniform priors in the parameters of the standard PMNS parameterisation privilege symmetries between the $\nu_{\mu(1)}$ and $\nu_{\tau(2)}$ flavour (mass) neutrino eigenstates and constructed a set of nine parameterisations that capture all such flavour and masss symmetries. We presented a method for applying priors induced by these parameterisations to Bayesian long-baseline neutrino oscillation analysis and discussed the additional uncertainties introduced by this process. Finally, we studied the changes to T2K's latest constraints arising from choosing the new priors. We found no significant alterations to the results on CP violation in neutrino oscillations; still, the current slight preference for the upper octant is sensitive to the choice of prior, and almost vanishes under some of these alternate constraints.

\begin{acknowledgements}
We thank the J-PARC staff for superb accelerator performance. We thank the CERN NA61/SHINE Collaboration for providing valuable particle production data. We acknowledge the support of MEXT,   JSPS KAKENHI (JP16H06288, JP18K03682, JP18H03701, JP18H05537, JP19J01119, JP19J22440, JP19J22258, JP20H00162, JP20H00149, JP20J20304) and bilateral programs (JPJSBP120204806, JPJSBP120209601), Japan; NSERC, the NRC, and CFI, Canada; the CEA and CNRS$/$IN2P3, France; the DFG (RO 3625$/$2), Germany; the NKFIH (NKFIH 137812 and TKP2021-NKTA-64), Hungary; the INFN, Italy; the Ministry of Education and Science(2023$/$WK$/$04) and the National Science Centre  (UMO-2018$/$30$/$E$/$ST2$/$00441   and  UMO-2022$/$46$/$E$/$ST2$/$00336 ), Poland; the RSF19-12-00325, RSF22-12-00358, Russia; MICINN (SEV-2016-0588, PID2019-107564GB-I00, PGC2018-099388-BI00, PID2020-114687GB-I00) Government of Andalucia (FQM160, SOMM17/6105/UGR) and the University of Tokyo ICRR's Inter-University Research Program FY2023 Ref. J1, and ERDF funds and CERCA program, Spain; the SNSF and SERI ($200021\_185012$, $200020\_188533$, 20FL21$\_$186178I), Switzerland; the STFC and UKRI, UK; and the DOE, USA. We also thank CERN for the UA1/NOMAD magnet, DESY for the HERA-B magnet mover system, the BC DRI Group, Prairie DRI Group, ACENET, SciNet, and CalculQuebec consortia in the Digital Research Alliance of Canada, GridPP and the Emerald High Performance Computing facility in the United Kingdom, and the CNRS/IN2P3 Computing Center in France. In addition, the participation of individual researchers and institutions has been further supported by funds from the ERC (FP7), “la Caixa” Foundation (ID 100010434, fellowship code LCF/BQ/IN17/11620050), the European Union’s Horizon 2020 Research and Innovation Programme under the Marie Sklodowska-Curie grant agreement numbers 713673 and 754496, and H2020 grant numbers RISE-GA822070-JENNIFER2 2020 and RISE-GA872549-SK2HK; the JSPS, Japan; the Royal Society, UK; French ANR grant number ANR-19-CE31-0001; the SNF Eccellenza grant number PCEFP2$\_$203261; and the DOE Early Career programme, USA. For the purposes of open access, the authors have applied a Creative Commons Attribution licence to any Author Accepted We thank the J-PARC staff for superb accelerator performance. We thank the CERN NA61/SHINE Collaboration for providing valuable particle production data. We acknowledge the support of MEXT,   JSPS KAKENHI (JP16H06288, JP18K03682, JP18H03701, JP18H05537, JP19J01119, JP19J22440, JP19J22258, JP20H00162, JP20H00149, JP20J20304) and bilateral programs(JPJSBP120204806, JPJSBP120209601), Japan; NSERC, the NRC, and CFI, Canada; the CEA and CNRS$/$IN2P3, France; the DFG (RO 3625$/$2), Germany; the NKFIH (NKFIH 137812 and TKP2021-NKTA-64), Hungary; the INFN, Italy; the Ministry of Education and Science(2023/WK/04) and the National Science Centre  (UMO-2018/30/E/ST2/00441   and  UMO-2022/46/E/ST2/00336 ), Poland; the RSF19-12-00325, RSF22-12-00358, Russia; MICINN (SEV-2016-0588, PID2019-107564GB-I00, PGC2018-099388-BI00, PID2020-114687GB-I00) Government of Andalucia (FQM160, SOMM17$/$6105$/$UGR) and the University of Tokyo ICRR's Inter-University Research Program FY2023 Ref. J1, and ERDF funds and CERCA program, Spain; the SNSF and SERI ($200021\_185012$, $200020\_188533$, 20FL21$\_$186178I), Switzerland; the STFC and UKRI, UK; and the DOE, USA. We also thank CERN for the UA1/NOMAD magnet, DESY for the HERA-B magnet mover system, the BC DRI Group, Prairie DRI Group, ACENET, SciNet, and CalculQuebec consortia in the Digital Research Alliance of Canada, GridPP and the Emerald High Performance Computing facility in the United Kingdom, and the CNRS/IN2P3 Computing Center in France. In addition, the participation of individual researchers and institutions has been further supported by funds from the ERC (FP7), “la Caixa” Foundation (ID 100010434, fellowship code LCF/BQ/IN17/11620050), the European Union’s Horizon 2020 Research and Innovation Programme under the Marie Sklodowska-Curie grant agreement numbers 713673 and 754496, and H2020 grant numbers RISE-GA822070-JENNIFER2 2020 and RISE-GA872549-SK2HK; the JSPS, Japan; the Royal Society, UK; French ANR grant number ANR-19-CE31-0001; the SNF Eccellenza grant number PCEFP$2\_203261$; and the DOE Early Career programme, USA. For the purposes of open access, the authors have applied a Creative Commons Attribution licence to any Author Accepted Manuscript version arising.
\end{acknowledgements}

\clearpage
\bibliographystyle{epj.bst}
\bibliography{references.bib}

\appendix
\clearpage

\section{Expanded forms of the PMNS matrix in 9 rotation parameterisations \label{sec:fullPMNS}}

Here we present the full form of the PMNS matrix under the 9 (6 Tait-Bryan and 3 Euler rotations) parameterisations considered in this analysis. Up to relabeling of the mixing angles and sign of the complex phase, six of these are equivalent to the parameterisations derived in~\cite{Denton} by considering different products of the matrices $R_{23}$, $\Gamma_\delta^\dagger R_{13}\Gamma_\delta,$ and $ R_{12}$. For example, $U_{\nu_\mu\nu_\tau / \nu_1\nu_3} \equiv P_{\mathbb{I}} \times U_{R} \times P_{-}$ is structurally equivalent to the Tait-Bryan rotation $R_{23}\times R_{12} \times \Gamma_\delta^\dagger \times R_{13}\times \Gamma_\delta$. Using the construction below, the intrinsic flavour and mass symmetries of each parameterisation become more obvious. The ordering of the parameterisations mirrors the label in Figure \ref{fig:1Dposterior}, and expression \ref{mutau12} is the canonical form. The correspondences with the parameterisations in ~\cite{Denton} may be identified by spotting the position of the single-angle element. The 9 matrices are generated by identifying  $X_1$ and $X_2$ with one of the three $3\times 3$ even permutation matrices (i.e. the action of the alternating group $A_3$):

\[
P_{\mathbb{I}} = \begin{pmatrix}
1 & 0 & 0 \\
0 & 1 & 0 \\
0 & 0 & 1 \\
\end{pmatrix}, \quad
P_+ = \begin{pmatrix}
0 & 1 & 0 \\
0 & 0 & 1 \\
1 & 0 & 0 \\
\end{pmatrix}, \quad
P_- = \begin{pmatrix}
0 & 0 & 1 \\
1 & 0 & 0 \\
0 & 1 & 0 \\
\end{pmatrix}.
\]

\begin{strip}
\begin{align}
U_{\nu_\mu\nu_\tau / \nu_1\nu_2} &= P_{\mathbb{I}} \times U_{R} \times P_{\mathbb{I}} = \begin{pmatrix}
      c_{12}c_{13} & s_{12}c_{13} &   s_{13}e^{-i \delta_{CP}}    \\
     -s_{12}c_{23} - c_{12}s_{23}s_{13}e^{i \delta_{CP}} &  c_{12}c_{23} - s_{12}s_{23}s_{13}e^{i \delta_{CP}} &        s_{23}c_{13}    \\
      s_{12}s_{23} - c_{12}c_{23}s_{13}e^{i \delta_{CP}} & -c_{12}s_{23} - s_{12}c_{23}s_{13}e^{i \delta_{CP}} &
      c_{23}c_{13}
    \end{pmatrix} \label{mutau12} \\[1.5em]
U_{\nu_\mu\nu_\tau / \nu_2\nu_3} &= P_{\mathbb{I}} \times U_{R} \times P_+ = \begin{pmatrix}
      s_{13}e^{-i \delta_{CP}} & c_{12}c_{13} & s_{12}c_{13}     \\
      s_{23}c_{13} & -s_{12}c_{23} - c_{12}s_{23}s_{13}e^{i \delta_{CP}} &  c_{12}c_{23} - s_{12}s_{23}s_{13}e^{i \delta_{CP}} \\
      c_{23}c_{13} & s_{12}s_{23} - c_{12}c_{23}s_{13}e^{i \delta_{CP}} & -c_{12}s_{23} - s_{12}c_{23}s_{13}e^{i \delta_{CP}} 
    \end{pmatrix} \label{mutau23} \\[1.5em]
U_{\nu_\mu\nu_\tau / \nu_1\nu_3} &= P_{\mathbb{I}} \times U_{R} \times P_- = \begin{pmatrix}
      c_{12}c_{13}  &   s_{13}e^{-i \delta_{CP}}  & s_{12}c_{13}  \\
     -s_{12}c_{23} - c_{12}s_{23}s_{13}e^{i \delta_{CP}}  & s_{23}c_{13} &  c_{12}c_{23} - s_{12}s_{23}s_{13}e^{i \delta_{CP}}   \\
      s_{12}s_{23} - c_{12}c_{23}s_{13}e^{i \delta_{CP}} & c_{23}c_{13} & -c_{12}s_{23} - s_{12}c_{23}s_{13}e^{i \delta_{CP}} 
    \end{pmatrix} \label{mutau13} \\[1.5em]
U_{\nu_e\nu_\tau / \nu_1\nu_2} &= P_- \times U_{R} \times P_{\mathbb{I}} = \begin{pmatrix}
     -s_{12}c_{23} - c_{12}s_{23}s_{13}e^{i \delta_{CP}} &  c_{12}c_{23} - s_{12}s_{23}s_{13}e^{i \delta_{CP}} &        s_{23}c_{13}    \\
      c_{12}c_{13} & s_{12}c_{13} &   s_{13}e^{-i \delta_{CP}}    \\
      s_{12}s_{23} - c_{12}c_{23}s_{13}e^{i \delta_{CP}} & -c_{12}s_{23} - s_{12}c_{23}s_{13}e^{i \delta_{CP}} &
      c_{23}c_{13}
    \end{pmatrix} \label{etau12} \\[1.5em] 
U_{\nu_e\nu_\tau / \nu_2\nu_3} &= P_+ \times U_{R} \times P_- = \begin{pmatrix}
      s_{23}c_{13} & -s_{12}c_{23} - c_{12}s_{23}s_{13}e^{i \delta_{CP}} &  c_{12}c_{23} - s_{12}s_{23}s_{13}e^{i \delta_{CP}}      \\
      s_{13}e^{-i \delta_{CP}} & c_{12}c_{13} & s_{12}c_{13}       \\
      c_{23}c_{13} & s_{12}s_{23} - c_{12}c_{23}s_{13}e^{i \delta_{CP}} & -c_{12}s_{23} - s_{12}c_{23}s_{13}e^{i \delta_{CP}} 
    \end{pmatrix} \\[1.5em]
U_{\nu_e\nu_\tau / \nu_1\nu_3} &= P_- \times U_{R} \times P_- = \begin{pmatrix}
     -s_{12}c_{23} - c_{12}s_{23}s_{13}e^{i \delta_{CP}} &        s_{23}c_{13}   &  c_{12}c_{23} - s_{12}s_{23}s_{13}e^{i \delta_{CP}}  \\
      c_{12}c_{13} &   s_{13}e^{-i \delta_{CP}}  & s_{12}c_{13}   \\
      s_{12}s_{23} - c_{12}c_{23}s_{13}e^{i\delta_{CP}}  & c_{23}c_{13}  & -c_{12}s_{23} - s_{12}c_{23}s_{13}e^{i \delta_{CP}}
    \end{pmatrix} \\[1.5em]
U_{\nu_e\nu_\mu / \nu_1\nu_2} &= P_+ \times U_{R} \times P_{\mathbb{I}} = \begin{pmatrix}
     -s_{12}c_{23} - c_{12}s_{23}s_{13}e^{i \delta_{CP}} &  c_{12}c_{23} - s_{12}s_{23}s_{13}e^{i \delta_{CP}} &        s_{23}c_{13}    \\
      s_{12}s_{23} - c_{12}c_{23}s_{13}e^{i \delta_{CP}} & -c_{12}s_{23} - s_{12}c_{23}s_{13}e^{i \delta_{CP}} &
      c_{23}c_{13} \\
      c_{12}c_{13} & s_{12}c_{13} &   s_{13}e^{-i \delta_{CP}}   
    \end{pmatrix} \label{emu12} \\[1.5em]
U_{\nu_e\nu_\mu / \nu_2\nu_3} &= P_+ \times U_{R} \times P_+ = \begin{pmatrix}
      s_{23}c_{13} & -s_{12}c_{23} - c_{12}s_{23}s_{13}e^{i \delta_{CP}} &  c_{12}c_{23} - s_{12}s_{23}s_{13}e^{i \delta_{CP}}   \\
      c_{23}c_{13} & s_{12}s_{23} - c_{12}c_{23}s_{13}e^{i \delta_{CP}} & -c_{12}s_{23} - s_{12}c_{23}s_{13}e^{i \delta_{CP}}  \\
      s_{13}e^{-i \delta_{CP}} & c_{12}c_{13} & s_{12}c_{13}   s_{13}e^{-i \delta_{CP}}   
    \end{pmatrix} \label{emu23} \\[1.5em] 
U_{\nu_e\nu_\mu / \nu_1\nu_3} &= P_+ \times U_{R} \times P_- = \begin{pmatrix}
     -s_{12}c_{23} - c_{12}s_{23}s_{13}e^{i \delta_{CP}} &   s_{23}c_{13}  &  c_{12}c_{23} - s_{12}s_{23}s_{13}e^{i \delta_{CP}}   \\
      s_{12}s_{23} - c_{12}c_{23}s_{13}e^{i \delta_{CP}} & c_{23}c_{13}  & -c_{12}s_{23} - s_{12}c_{23}s_{13}e^{i \delta_{CP}} \\
      c_{12}c_{13} &   s_{13}e^{-i \delta_{CP}}  & s_{12}c_{13}   
    \end{pmatrix} \label{emu13}
\end{align}
\end{strip}

\section{Fake data results \label{sec:fakeResults}}
To test whether the main results in section \ref{sec:results} are due to the robustness of T2K's constraining power or due to the particular shape of the likelihood in the favoured area of parameter space, we run the same analysis on chains generated from two Asimov datasets~\cite{Asimov} at points Asimov A and Asimov B of the phase space (defined in table \ref{tab:asimov}). Asimov A is chosen to recreate posteriors similar to the data fit, and Asimov B is chosen to represent a scenario with no CP violation and true lower octant.

\begin{figure*}
    \centering
    \includegraphics[width=0.9\textwidth]{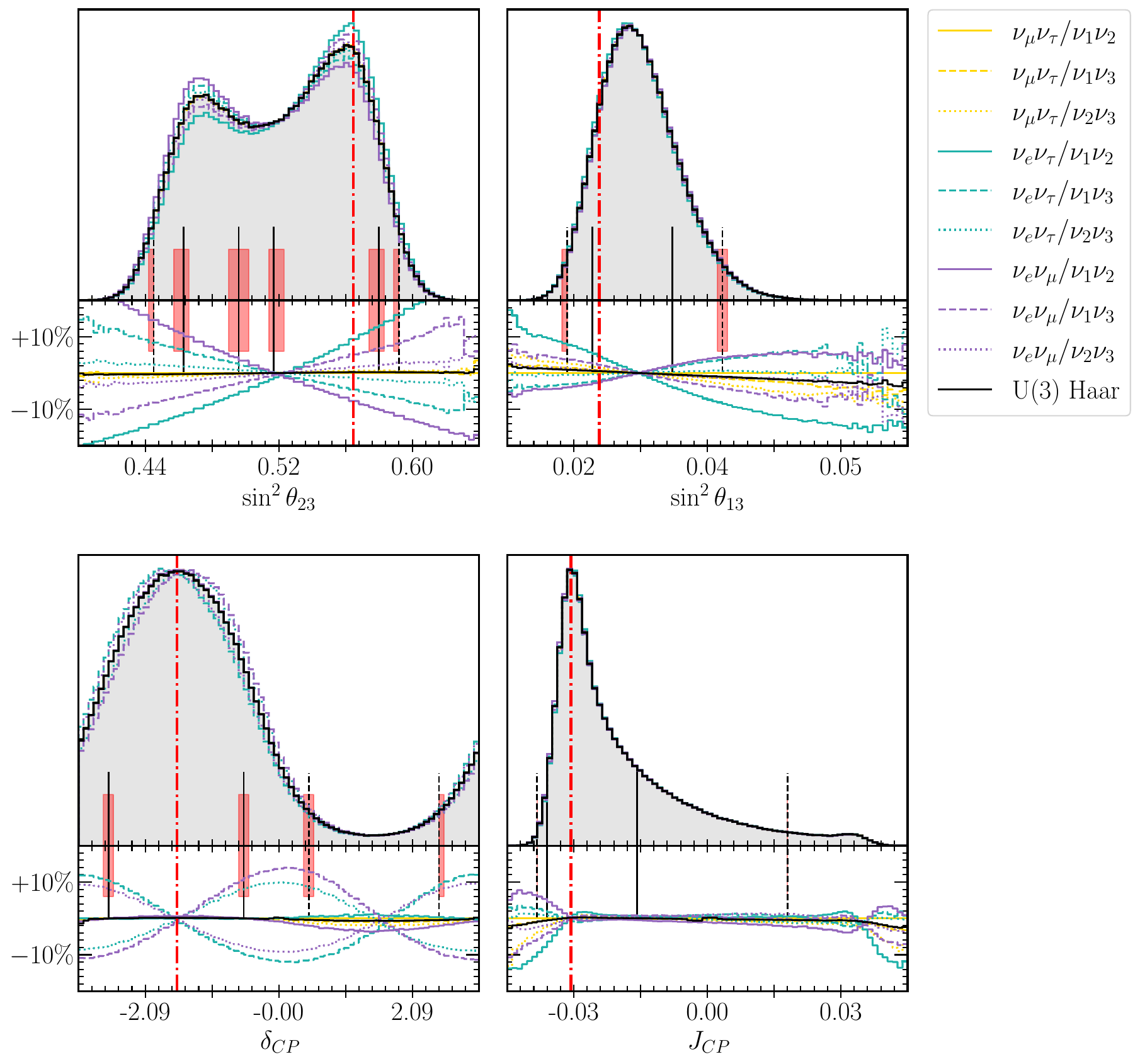}
    \caption{1D marginalised posterior over the standard parameters of a T2K Asimov A fit, reweighted under the 9 flavour/mass symmetry and Haar priors. The black line corresponds to the original posterior generated using the PDG prior, and the vertical lines mark the $1\sigma$ (filled) and $2\sigma$ (dashed) credible regions. The bottom plot shows the fractional bin change from the standard prior. The Asimov point is marked with a red line.}
    \label{fig:1DposteriorAsimovA}
\end{figure*}

\begin{figure*}
    \centering
    \includegraphics[width=0.9\textwidth]{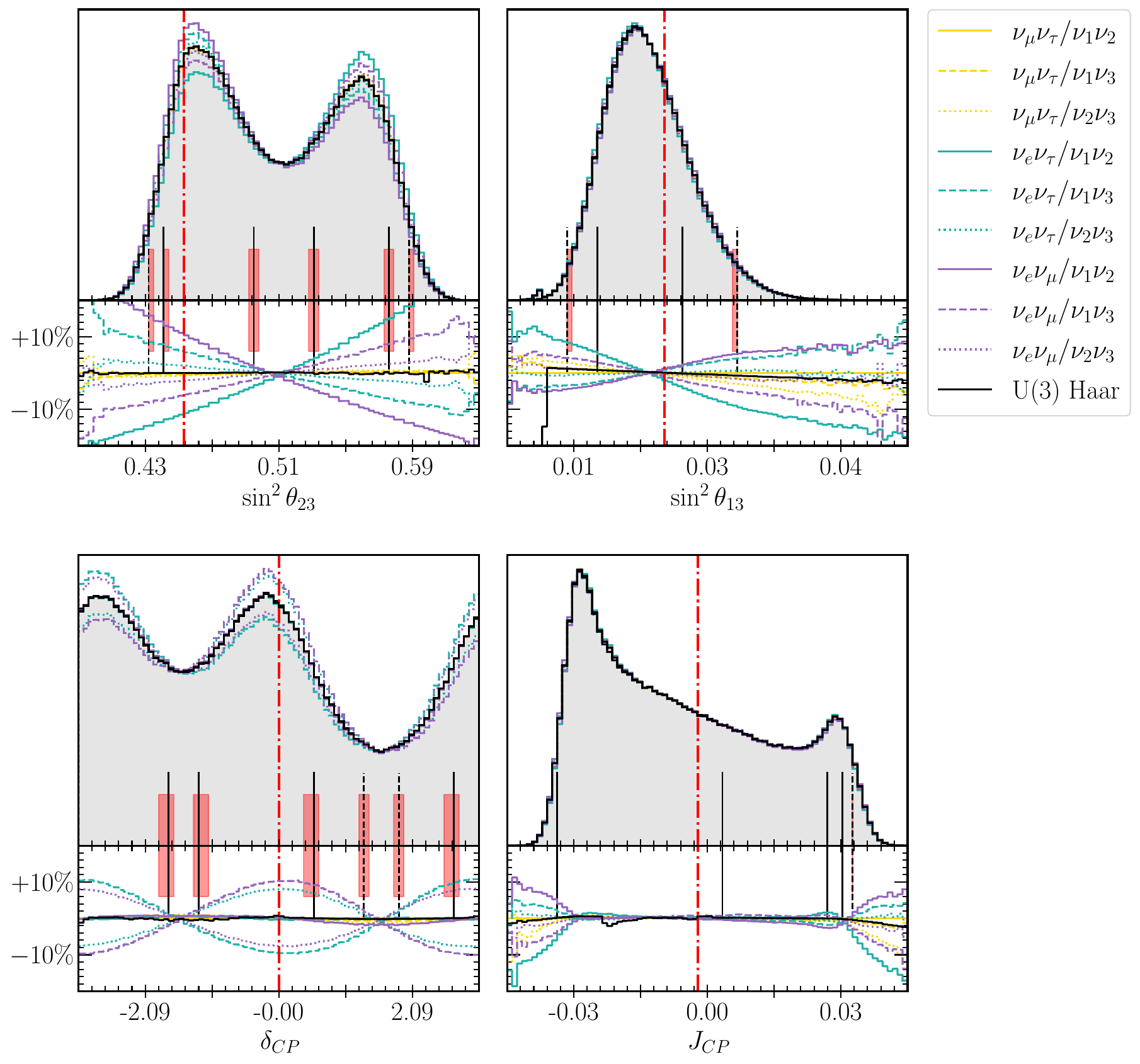}
    \caption{1D marginalised posterior over the standard parameters of a T2K Asimov B fit, reweighted under the 9 flavour/mass symmetry and Haar priors. The black line corresponds to the original posterior generated using the PDG prior, and the vertical lines mark the $1\sigma$ (filled) and $2\sigma$ (dashed) credible regions. The bottom plot shows the fractional bin change from the standard prior. The Asimov point is marked with a red line.}
    \label{fig:1DposteriorAsimovB}
\end{figure*}

Figures \ref{fig:1DposteriorAsimovA} and \ref{fig:1DposteriorAsimovB} show the 1D marginalised posteriors for Asimov points A and B, and reweighted with the Tait-Bryan priors. The results are consistent with the conclusions of section \ref{sec:results}: the $\theta_{23}$ octant preference vanishes for some priors but the constraints remain largely the same. Although the posteriors for $\delta_{CP}$ vary substantially (Asimov B experiences a shift of the highest posterior density from 0 to $\pm\pi$), the constraints on the amount of CP violation as given by the Jarlskog invariant show sub-percent change within the $2\sigma$ range. This falls in line with the fact that T2K has good sensitivity to the observable $J_{CP}$, which is a more robust measure of CP-violation than a prior-dependent extraction to $\delta_{CP}$.

\begin{table}[h]
\centering
\begin{tabular}{lll}
\hline
Parameter                      & Asimov A value         & Asimov B value         \\ \hline
$\sin^2\theta_{12}^\text{PDG}$ & $0.307$                & $0.307$                \\
$\sin^2\theta_{23}^\text{PDG}$ & $0.561$                & $0.45$                 \\
$\sin^2\theta_{13}^\text{PDG}$ & $0.022$                & $0.022$                \\
$\delta_{CP}^\text{PDG}$       & $-1.601$               & $0$                    \\
$\Delta m^2_{21} $             & $2.494 \times 10^{-3}$ eV$^2$ & $2.494 \times 10^{-3}$ eV$^2$ \\
$\Delta m^2_{32}$              & $7.53 \times 10^{-5}$ eV$^2$  & $7.53 \times 10^{-5}$ eV$^2$ 
\end{tabular}
\caption{Parameter values for Asimov fits in the T2K experiment. }
\label{tab:asimov}
\end{table}


\section{Data release}
This study further analyses the results reported in~\cite{T2K:2025ana}, which provides the associated data.

\end{document}